\newcommand{\argmin}{\operatornamewithlimits{argmin}}
\newcommand{\off}{\operatornamewithlimits{off}}

\documentclass[10pt,conference]{IEEEtran}
\usepackage{times,amsmath,epsfig}
\usepackage[T1]{fontenc}
\usepackage[latin9]{inputenc}
\usepackage{listings}
\usepackage{float}
\usepackage{textcomp}
\usepackage{graphicx}
\floatstyle{ruled}
\newfloat{algorithm}{tbp}{loa}
\floatname{algorithm}{Algorithm}

\usepackage{url}
\title{On Joint Diagonalisation \\for Dynamic Network Analysis}
\author{%
{Damien Fay {\small $~^{\#1}$}, J\'er\^ome Kunegis{\small $~^{\&2}$}, Eiko Yoneki{\small $~^{\$3}$}}%
\vspace{1.6mm}\\
\fontsize{10}{10}\selectfont\itshape
$~^{\#}$University of Cork, Ireland\\
$~^{\&}$University of Koblenz--Landau, Germany\\
$~^{\$}$University of Cambridge, United Kingdom\\
\fontsize{9}{9}\selectfont\ttfamily\upshape
\vspace{1.6mm}\\
$~^{1}$damien.fay@cl.cam.ac.uk\\
$~^{2}$kunegis@uni-koblenz.de\\
$~^{3}$eiko.yoneki@cl.cam.ac.uk }

\begin{document}
\maketitle
\begin{abstract}
Joint diagonalisation (JD) is a technique used to estimate an average eigenspace of a set of matrices. Whilst
it has been used successfully in many areas to track the evolution of systems via their eigenvectors; its
application in network analysis is novel. The key focus in this paper is the use of JD on matrices of spanning
trees of a network.  This is especially useful in the case of real-world contact networks in which a single
underlying static graph does not exist. The average eigenspace may be used to construct a graph which
represents the `average spanning tree' of the network or a representation of the most common propagation paths.
We then examine the distribution of deviations from the average and find that this distribution in real-world
contact networks is multi-modal; thus indicating several \emph{modes} in the underlying network. These modes
are identified and are found to correspond to particular times. Thus JD may be used to decompose the behaviour,
in time, of contact networks and produce average static graphs for each time.  This may be viewed as a mixture
between a dynamic and static graph approach to contact network analysis.
\end{abstract}

% NOTE keywords are not used for conference papers so do not populate them

\begin{figure*}
 \centering
   \includegraphics[width=9.5cm]{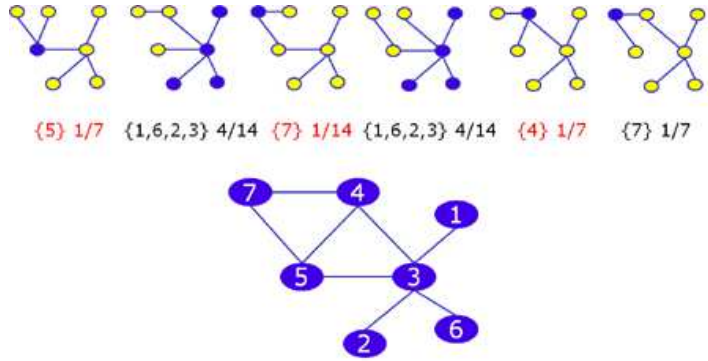}
 \caption{A simple graph and its 6 spanning trees. (The numbers represent the root nodes and probability of observing the tree ex: {1,6,2,3} are the root nodes for the third tree and this tree is observed with probability 4/14)\label{f:simple_graph_ex}}
\end{figure*}
%%%%%%%%%%%%%%%%%%%%%%%%%%%%%%%%%%%%%%%%%%%
%%%%%%%%%%%%%%%%%%%%%%%%%%%%%%%%%%%%%%%%%%%
%%%%%%%%%%%%%%%%%%%%%%%%%%%%%%%%%%%%%%%%%%%
\section{Introduction}
%%%%%%%%%%%%%%%%%%%%%%%%%%%%%%%%%%%%%%%%%%%
%%%%%%%%%%%%%%%%%%%%%%%%%%%%%%%%%%%%%%%%%%%
%%%%%%%%%%%%%%%%%%%%%%%%%%%%%%%%%%%%%%%%%%%
\label{sec:intro} Understanding the dynamic structure of contact networks is critical for designing dynamic
routing algorithms~\cite{Ioannidis}, epidemic spreading ~\cite{Riolo} and message passing
algorithms~\cite{hui:bubblerap}.

Time dependent networks are characterised by time dependant paths which are characterised by the order in which
the paths occur. For example, a path between 3 nodes $A\rightarrow B \rightarrow C$ does not imply a reverse
path exists; $A\rightarrow B \rightarrow C$ provides no information about how $C$ may communicate with $A$.
However, in many applications a static graph is constructed which represents typically the proportion of time a
link was seen between two nodes. These static graphs often lose the time information which is critical in
contact networks. However, at a \emph{specific time} and from a \emph{specific node} there is a single static
network representing the paths between the root node and the rest of the network. JD is used in this paper to
look for commonalities in these specific static networks and to provide an \emph{average} static network where
appropriate. That is, we are looking for \emph{modes} of operation in contact networks.

Joint Diagonalisation (JD) is a technique that is used to track the changes in eigenspace (i.e. eigenvectors
and eigenvalues) of a system (see Section~\ref{sec:theory} for examples).  Eigenvectors and eigenvalues play
an important role in static network/graph analysis as they can be used to determine the centrality of nodes;
communities and settling times among other things~\cite{network_analysis}. However, to the best of our
knowledge tracking eigenspace evolution has not been applied to contact/time dependent networks previously.
This paper examines the use of JD in network analysis.

%%%%%%%%%%%%%%%%%%%%%%%%%%%%%%%%%%%%%%%%%%%
%%%%%%%%%%%%%%%%%%%%%%%%%%%%%%%%%%%%%%%%%%%
%%%%%%%%%%%%%%%%%%%%%%%%%%%%%%%%%%%%%%%%%%%
\section{Related work}
\label{sec:related}
%%%%%%%%%%%%%%%%%%%%%%%%%%%%%%%%%%%%%%%%%%%
%%%%%%%%%%%%%%%%%%%%%%%%%%%%%%%%%%%%%%%%%%%
%%%%%%%%%%%%%%%%%%%%%%%%%%%%%%%%%%%%%%%%%%%

Joint diagonalisation has been used in many applications where the evolution of a system can be tracked
smoothly via its eigenspace. For example, Macagnano et al.~\cite{Macagnano2} present an algorithm for
localisation of multiple objects given partial location information. As time evolves the location of the
objects changes smoothly which may be seen through the evolution of the eigenvectors of a distance matrix.
Other examples include blind beam forming~\cite{Cardoso_beam} and blind source separation ~\cite{Wenwu}.

Sun et al.~\cite{b515} use tensor analysis to examine time dependent networks. A tensor is multi-dimensional
matrix (for example a set of adjacency matrices) which are essentially reduced using PCA to a \emph{core}
tensor. This technique is similar in spirit to that presented here, the difference is that we are looking to
reduce a set of spanning trees representing propagation through a network; propagation information being
preserved.

Scellato et al.~\cite{Scellato} examine the different characteristics of contact networks as they evolve over
time. However, the analysis there is based on forming static graphs by amalgamating all links seen in an
interval of time. This may introduce connections which in fact are unordered. Graph measures (e.g.\ the clustering
coefficient) are then measured from these graphs and a time series analysis of these follows. In contrast, here
the amalgamation is dependent of the contact network itself.

Riolo et al.~\cite{Riolo} investigate time dependent epidemic networks with a view to constructing
\emph{transmission graphs}, directed graphs which indicate the direction of transmission of a disease through a
network. While the aim in this paper is similar, the methodology used is significantly different as they
examine one time infections in real networks.

%%%%%%%%%%%%%%%%%%%%%%%%%%%%%%%%%%%%%%%%%%%
%%%%%%%%%%%%%%%%%%%%%%%%%%%%%%%%%%%%%%%%%%%
%%%%%%%%%%%%%%%%%%%%%%%%%%%%%%%%%%%%%%%%%%%
\section{Theoretical Background}
\label{sec:theory}
%%%%%%%%%%%%%%%%%%%%%%%%%%%%%%%%%%%%%%%%%%%
%%%%%%%%%%%%%%%%%%%%%%%%%%%%%%%%%%%%%%%%%%%
%%%%%%%%%%%%%%%%%%%%%%%%%%%%%%%%%%%%%%%%%%%
We begin by defining snowball sampling which consists of selecting a root node randomly in the network with
uniform probability and performing a \emph{Breath First Search} (BFS) from this node (i.e.\ determining a set
of shortest paths from the source node to every other node in the network). This produces a spanning tree, $H$,
where $H$ is a subset of the original graph $G(V,E)$, where $V$ and $E$ denote the vertex and edge sets
respectively, and $|V| = N$ denotes the number of nodes. We call the starting node the \emph{observer} or root
and $H$, \emph{the sample}. Figure~\ref{f:simple_graph_ex} shows a simple graph which will be used for
demonstration purposes. In the first sample node 5 is selected at random and a shortest path first search
results in the first tree in Figure~\ref{f:simple_graph_ex}. In this simple graph there are 6 spanning trees
shown in Figure~\ref{f:simple_graph_ex}. Note that the distribution of spanning trees in this network are not
uniform but biased. That is, traffic generated uniformly from each node results in non-uniform percolation
across the network.

% Many centrality techniques (as pointed out in Section~\ref{sec:related}) assume that each of these spanning trees is equally likely. However, this is not the case. For example the last spanning trees in Figure~\ref{f:simple_graph_ex} can only be observed if node 7 is chosen as the root node. As the probability of choosing node 7 is $\frac{1}{7}$ and there are two possible spanning trees the probability of observing each spanning tree is $\frac{1}{14} \ne \frac{1}{6}$ and so the spanning trees are not equally likely. The key point here is that \emph{the sampling mechanism introduces a bias on the spanning tree probabilities}.

%\begin{figure}
% \centering
%   \includegraphics[width=3.4in]{weird.eps}
% \caption{A simple graph and its 6 spanning trees. (The numbers represent the root nodes and probability of observing the tree ex: {1,6,2,3} are the root nodes for the third tree and this tree is observed with probability 4/14)\label{f:simple_graph_ex}}
%\end{figure}

Next we develop a centrality measure which is based on standard eigenvector centrality. Eigenvector centrality~\cite{Opsahl} is defined by letting the centrality of node $i$ equal the average of the centrality of all nodes connected to it:
\begin{equation}\label{e:centrality}
x_i = \frac{1}{\lambda} \sum_{j=1}^{N} A_{i,j} x_j
\end{equation}
where $A_{i,j}$ is element $i,j$ of the (possibly weighted) adjacency matrix and $\lambda$ is the largest eigenvalue of $A_{i,j}$ as can be seen by rewriting Equation~\ref{e:centrality} in matrix notation as:
\begin{equation}\label{e:evect}
X = \frac{1}{\lambda} AX
\end{equation}
The eigenvector corresponding to $\lambda_{max}$ gives the eigenvector centrality of node $i$.

Given $M$ samples of a network, $H_1 \ldots H_M$, the question now arises; how can these be combined to give a
matrix that reflects the sampling bias. We propose using a method known as joint diagonalisation which produces
an average eigenspace of the samples. Specifically, we seek an orthogonal matrix such that:
\begin{equation}\label{e:joint1}
H_i = UC_iU^T  \textnormal{    } \forall{i}
\end{equation}
If $U$ corresponds to the eigenvectors of $H_i$ then $C_i$ is diagonal however no matrix $U$ exists in which all $C_i$ are diagonal (except for the trivial case in which all $H_i$ are equal). Joint diagonalisation seeks average eigenvectors $\bar{U}$ such that the sum of squares of the off diagonal elements of $C_i$ are minimised. Specifically:
\begin{equation}\label{e:joint2}
\bar{U} = \argmin_{U} \normalfont{\off}_2(\sum_{j=1}^{M} C_i)
\end{equation}
where $\normalfont{\off}_2$ is the sum of the off diagonal elements squared, called the \emph{deviation} of $H_i$ from $\bar{H}$, $\delta_i$ :
\begin{equation}\label{e:joint3}
\delta_i = \normalfont{\off}_2(C_i) = \sum_{k\ne j } |C_i^{k,j}|^2
\end{equation}
where $C_i^{k,j}$ is the $k^{th}$ row and $j^{th}$ column of $C_i$.
As shown in~\cite{b350} and~\cite{Cardoso} Equation~\ref{e:joint2} may be minimised efficiently by a sequence of Givens rotations; convergence and stability properties are proven in~\cite{Bunse}.

Given the average eigenstructure of the sample matrices an average sampling matrix may be constructed from the eigenvector decomposition as:
\begin{equation}\label{e:joint4}
\bar{H}  = U \bar{C} U^T
\end{equation}
Where $\bar{H}$ is a matrix in which the entries represent the average weight of the links as observed by the samples in the network (in a least squares sense) and $\bar{C}$ is the average of diagonals of $H_i$ projected onto $\bar{U}$; i.e.\ the average eigenvalues. A sample based centrality may then be constructed from $\bar{H}$ using the standard eigenvector centrality; i.e.\ by using the eigenvector of $\bar{H}$ corresponding to the maximum eigenvalue.

\begin{figure}[t]
 \centering
   \includegraphics[width=7cm]{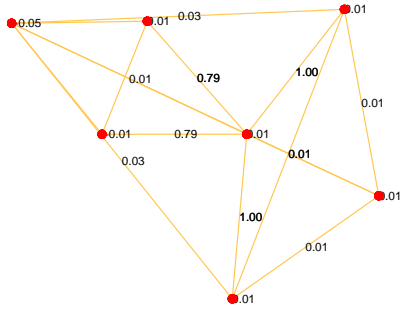}
 \caption{The average sampling graph corresponding to $\bar{H}$. \label{f:sampling_graph}}
\end{figure}
\subsection{Simple examples.}
Using the graph shown in Figure~\ref{f:simple_graph_ex}, 100 samples are taken by randomly choosing a root node and constructing a spanning tree from each. These are then jointly diagonalised producing the eigenvectors shown in Table~\ref{t:table1}. The sampling centrality is the first eigenvector (column 1).
\begin{figure}[b]
 \centering
   \includegraphics[width=6cm]{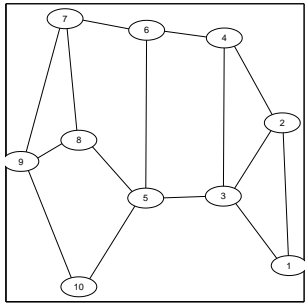}
 \caption{The simple graph used in example 2. \label{f:ex2full}}
\end{figure}

\begin{table}[h!]
\newcommand\T{\rule{0pt}{2.6ex}}
\begin{center}
{
\scriptsize
\caption{The average eigenvector, $U$ from the graph in Figure~\ref{f:simple_graph_ex}. \label{opt-table}}
\begin{tabular}{|c|c|c|c|c|c|c|}
 \hline
      1 &     2 &     3 &     4 &     5 &     6 &     7 \\ \hline
 0.3095 &  0.74 & -0.34 & -0.29 &  0.13 & -0.35 &  0.14 \\
 0.6780 & -0.00 &  0.70 & -0.19 & -0.09 &  0.00 & -0.09 \\
 0.3370 &  0.00 & -0.26 &  0.47 & -0.73 &  0.00 &  0.26 \\
 0.3357 &  0.00 & -0.26 &  0.45 &  0.28 &  0.00 & -0.73 \\
 0.3095 & -0.07 & -0.34 & -0.29 &  0.13 &  0.81 &  0.14 \\
 0.1634 & -0.00 &  0.14 &  0.54 &  0.57 &  0.00 &  0.58 \\
 \hline
\end{tabular}\label{t:table1}
}
\end{center}
\end{table}

It is instructive to view the eigenvector reconstruction of $\bar{H}$ which is formed from the eigenvectors in Table~\ref{t:table1} via Equation~\ref{e:joint4}. $\bar{H}$  is drawn in Figure~\ref{f:sampling_graph}. Note that $\bar{H}$ is a complete weighted graph; weights assigned to non-existent links are low and are a consequence of taking an average of many graphs. The edge between nodes 3 and 4 has a weight of $0.7857 \approx \frac{11}{14}$, i.e.\ the proportion of trees that use that link (see Figure~\ref{f:simple_graph_ex}).  $\bar{H}$ represents the sample biased weight of this link; a nice result. There are two reasons why these numbers are not exactly the same; the first is that, as always, there is a slight error introduced when using empirical sampling. The second is that the $\bar{H}$ is based on the average eigenspace of a set of sampling trees; this is not the same as simply taking the proportion of times a link has been observed.

\begin{figure}[t]
 \centering
   \includegraphics[width=7cm]{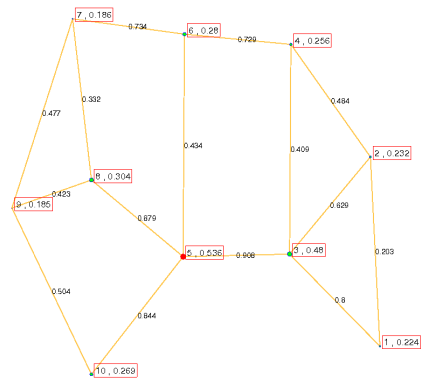}
 \caption{$\bar{H}$ for example 2 (Note: the red box contains the node number and the sampling centrality). \label{f:ex2b}}
\end{figure}

% \begin{figure}
%  \centering
%    \includegraphics[width=3.4in]{examplee1.eps}
%  \caption{$\bar{H}$ for the simple graph using routing which prefers to go via upper bridge.  \label{f:ex2c}}
% \end{figure}

\begin{figure}[h]
 \centering
   \includegraphics[width=7cm]{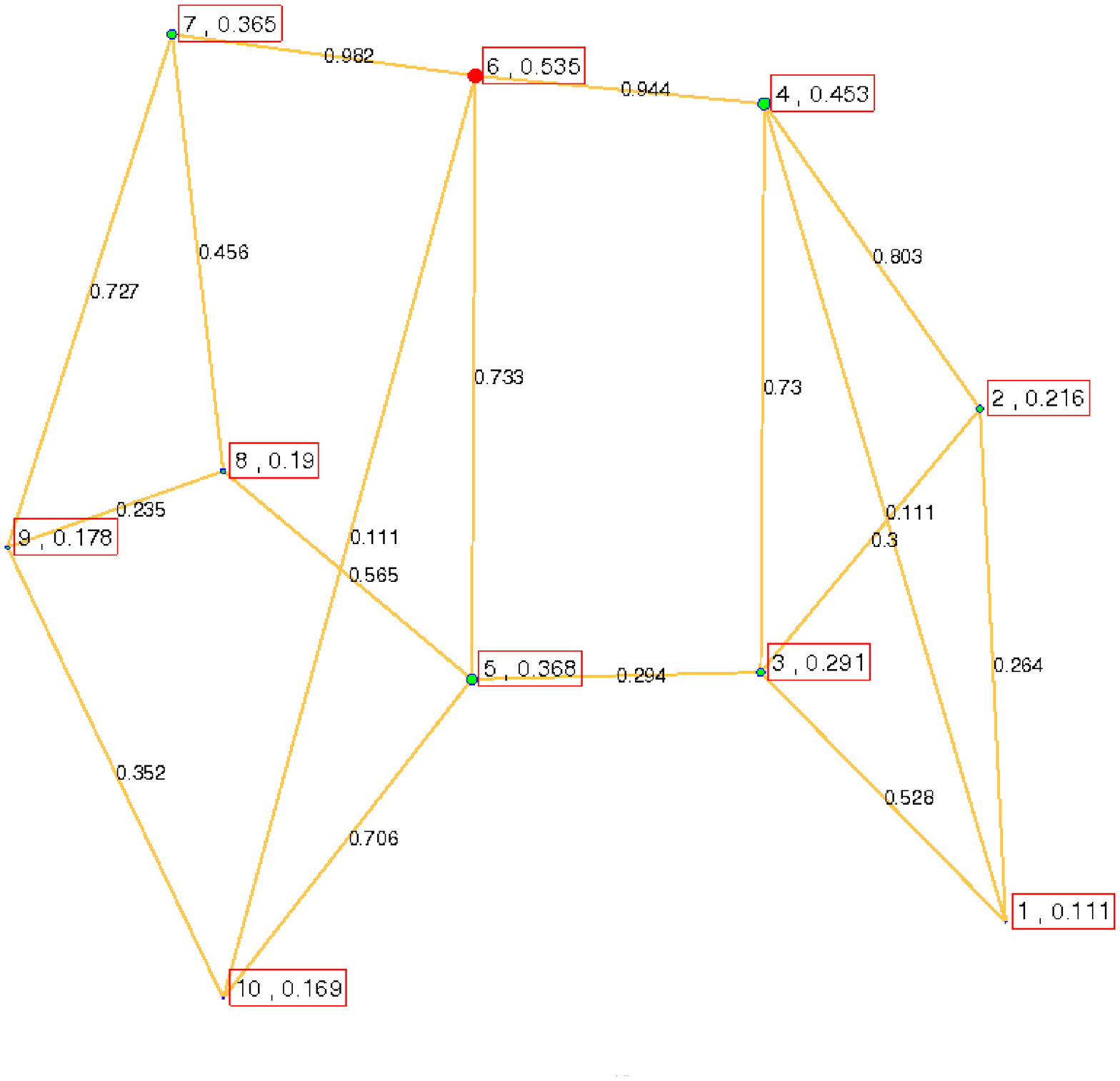}
 \caption{$\bar{H}$ for example 2 with a preference for routes 6$\leftrightarrow$4. \label{f:ex2d}}
\end{figure}
% \begin{table*}[htp]
% \newcommand\T{\rule{0pt}{2.6ex}}
% \begin{center}
% {
% \scriptsize
% \caption{$\bar{H}$ also shown in Figure~\ref{f:sampling_graph}. \label{opt-table2}}
% \begin{tabular}{|c|c|c|c|c|c|c|}
%  \hline
%      1 &     2 &    3 &     4 &     5 &     6 &     7 \\ \hline
%   0.01 &  0.01 & 1.00 & -0.04 & -0.03 &  0.01 &  0.03 \\
%   0.01 &  0.01 & 1.00 & -0.04 & -0.03 &  0.01 &  0.03 \\
%   1.00 &  1.00 & 0.01 &  0.79 &  0.79 &  1.00 &  0.01 \\
%  -0.04 & -0.04 & 0.79 &  0.01 &  0.43 & -0.04 &  0.49 \\
%  -0.03 & -0.03 & 0.79 &  0.43 & -0.01 & -0.03 &  0.49 \\
%   0.01 &  0.01 & 1.00 & -0.04 & -0.03 &  0.01 &  0.03 \\
%   0.03 &  0.03 & 0.01 &  0.49 &  0.49 &  0.03 & -0.05 \\
%  \hline
% \end{tabular}\label{t:table2}
% }
% \end{center}
% \end{table*}

The second example deals with a preferred route. Figure ~\ref{f:ex2full} shows a graph with a bridge formed from nodes 6$\leftrightarrow$4 and 5$\leftrightarrow$3 ; these nodes are critical in joining the two parts of the graph.

Figure~\ref{f:ex2b} shows the weighted graph created by sampling as above\footnote{Links with low weights are removed for clarity.}. The second stage involves creating a preferred route by removing 80\% of trees that use the link between 5$\leftrightarrow$3. The resultant average graph is shown in Figure~\ref{f:ex2d}. As can be seen the weight attached to link 5$\leftrightarrow$3 is greatly reduced (from 0.9 to 0.2).
%
% \begin{figure}
%  \centering
%    \includegraphics[width=3.4in]{examplee4.eps}
%  \caption{$\bar{H}$ for the simple graph using shortest path routing. \label{f:ex2a}}
% \end{figure}

Thus far we have only dealt with samples taken on static graphs. However, joint diagonalisation is particularly well suited to contact networks. In these networks there is no underlying static graph as such, but rather a set of contacts that are time dependent. By flooding these networks spanning trees may be formed and combined by the use of JD. The next section details these real-world data sets.
%%%%%%%%%%%%%%%%%%%%%%%%%%%%%%%%%%%%%%%%%%%
%%%%%%%%%%%%%%%%%%%%%%%%%%%%%%%%%%%%%%%%%%%
%%%%%%%%%%%%%%%%%%%%%%%%%%%%%%%%%%%%%%%%%%%
\section{Data set details}\label{sec:data}
%%%%%%%%%%%%%%%%%%%%%%%%%%%%%%%%%%%%%%%%%%%
%%%%%%%%%%%%%%%%%%%%%%%%%%%%%%%%%%%%%%%%%%%
%%%%%%%%%%%%%%%%%%%%%%%%%%%%%%%%%%%%%%%%%%%

In this paper, we use four experimental datasets gathered by the Haggle Project {\cite{haggle}},
referred to as \emph{Cambridge}, \emph{Infocom06}; one dataset from the MIT Reality Mining
Project~\cite{realityMining}, referred to as \emph{MIT}. Previously, the characteristics of these
datasets such as inter-contact and contact distribution have been explored in several
studies~\cite{asonam}, to which we refer the reader for further background information. These three
datasets cover a rich diversity of environments, ranging from a quiet university town
(\emph{Cambridge}), with an experimental period from a few days (\emph{Infocom06}) to one
month (\emph{MIT}).
\begin{table}[h]
 \begin{center}
\resizebox{8cm}{!} {
 \begin{tabular}{|c|c|c|c|}
 \hline Experimental data set
 &Cambridge & Infocom06 & MIT\\
 \hline
 Device
 &iMote & iMote & Phone\\
 Network type
 &Bluetooth & Bluetooth & Bluetooth \\
Duration (days)
 & 11   &  3  & 246\\
 Granularity (seconds)
 & 600  &  120 & 300 \\
 Number of Devices
  &36   &  78 & 97\\
 Number of contacts
 &10,873  &  191,336  & 54,667\\
 Average \# Contacts/pair/day
 & 0.345  &  6.7  & 0.024\\
\hline
 \end{tabular}
}
 \caption{\footnotesize{Characteristics of experimental data sets}}
 \label{table:datasets}
 \end{center}
 \end{table}

\begin{itemize}
\item In \emph{Cambridge}, the iMotes were distributed mainly to two
groups of students from University of Cambridge Computer Laboratory,
specifically undergraduate year1 and year2 students, and also some
PhD and Masters students. This dataset covers 11 days.

\item In \emph{Infocom06}, the trace contains 78 participants.
Among 78 participants, 34 form 4 subgroups by academic affiliations.

\item In \emph{MIT}, 100 smart phones were deployed
to students and staff at MIT over a period of 9 months. These phones
were running software that logged contacts with other Bluetooth
enabled devices by doing Bluetooth device discovery every five
minutes. 1 month of data is used here to maintain the consistency of users.
\end{itemize}

The three experiments are summarised in Table~\ref{table:datasets} and the trace data can be
downloaded at CRAWDAD database~\cite{crawdad}.

%%%%%%%%%%%%%%%%%%%%%%%%%%%%%%%%%%%%%%%%%%%
%%%%%%%%%%%%%%%%%%%%%%%%%%%%%%%%%%%%%%%%%%%
%%%%%%%%%%%%%%%%%%%%%%%%%%%%%%%%%%%%%%%%%%%
\section{Results}
\label{sec:results}
%%%%%%%%%%%%%%%%%%%%%%%%%%%%%%%%%%%%%%%%%%%
%%%%%%%%%%%%%%%%%%%%%%%%%%%%%%%%%%%%%%%%%%%
%%%%%%%%%%%%%%%%%%%%%%%%%%%%%%%%%%%%%%%%%%%
The results below examine the 3 data sets separately highlighting the features of each. Finally a synthetic
contact network with known characteristics is constructed and JD used to extract these characteristics.

\begin{figure}[htb]
 \centering
   \includegraphics[width=8cm]{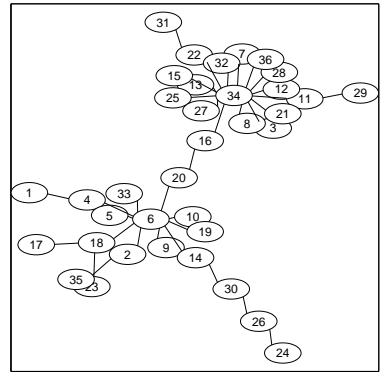}
 \caption{A typical flooding tree; as seen from node 20. \label{f:camb2}}
\end{figure}
\begin{figure}[b]
 \centering
   \includegraphics[width=6cm]{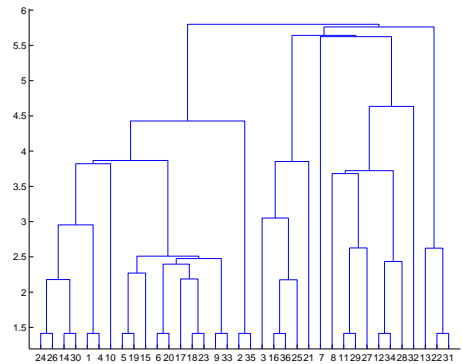}
 \caption{Community based on Fiedler clustering (Cambridge data set). \label{f:camb_grouped}}
\end{figure}
\begin{figure}[tb]
 \centering
   \includegraphics[width=8cm]{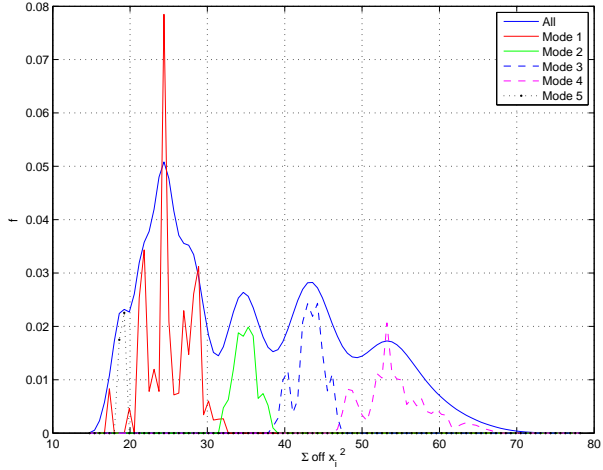}
 \caption{Distribution of $\delta_i$ (Cambridge data set; kernel smoothing is employed for the overall average.). \label{f:camb_dist}}
\end{figure}

\begin{figure}[htb]
 \centering
   \includegraphics[width=8cm]{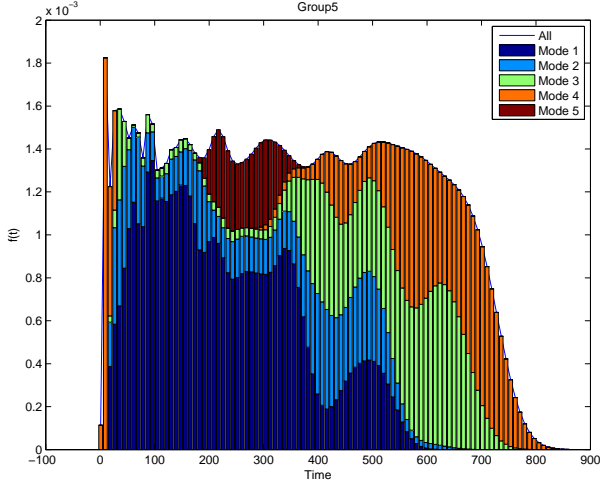}
 \caption{Distribution of times by mode. \label{f:camb_grouped_times}}
\end{figure}

Next the distribution of the sample start times is examined Figure~\ref{f:camb_grouped_times}. As can be seen the 5 modes correspond to different times in the data set. Modes 1 and 5 cover the first half of the data while modes 2 then 3 and then 4 become dominant in that succession.

\subsection{Cambridge data}
\label{sec:camb_data}
Figure~\ref{f:camb2} shows a typical sampling tree for the Cambridge data set. Node 20 initiates a message and it is passed around the contact network; first to nodes 16 and 6 and from there to the rest of the network. For this experiment ten thousand such trees are generated\footnote{A large sample size is used here to negate random effects. However, similar results are found for much smaller sample sizes.} with the messages starting at a random times and from a random node (uniformly distributed). These are then combined using Joint Diagonalization to form $\bar{H}$.

\begin{figure*}[htb]
\begin{center}$
\begin{array}{c c c}
\includegraphics[width=4.2cm]{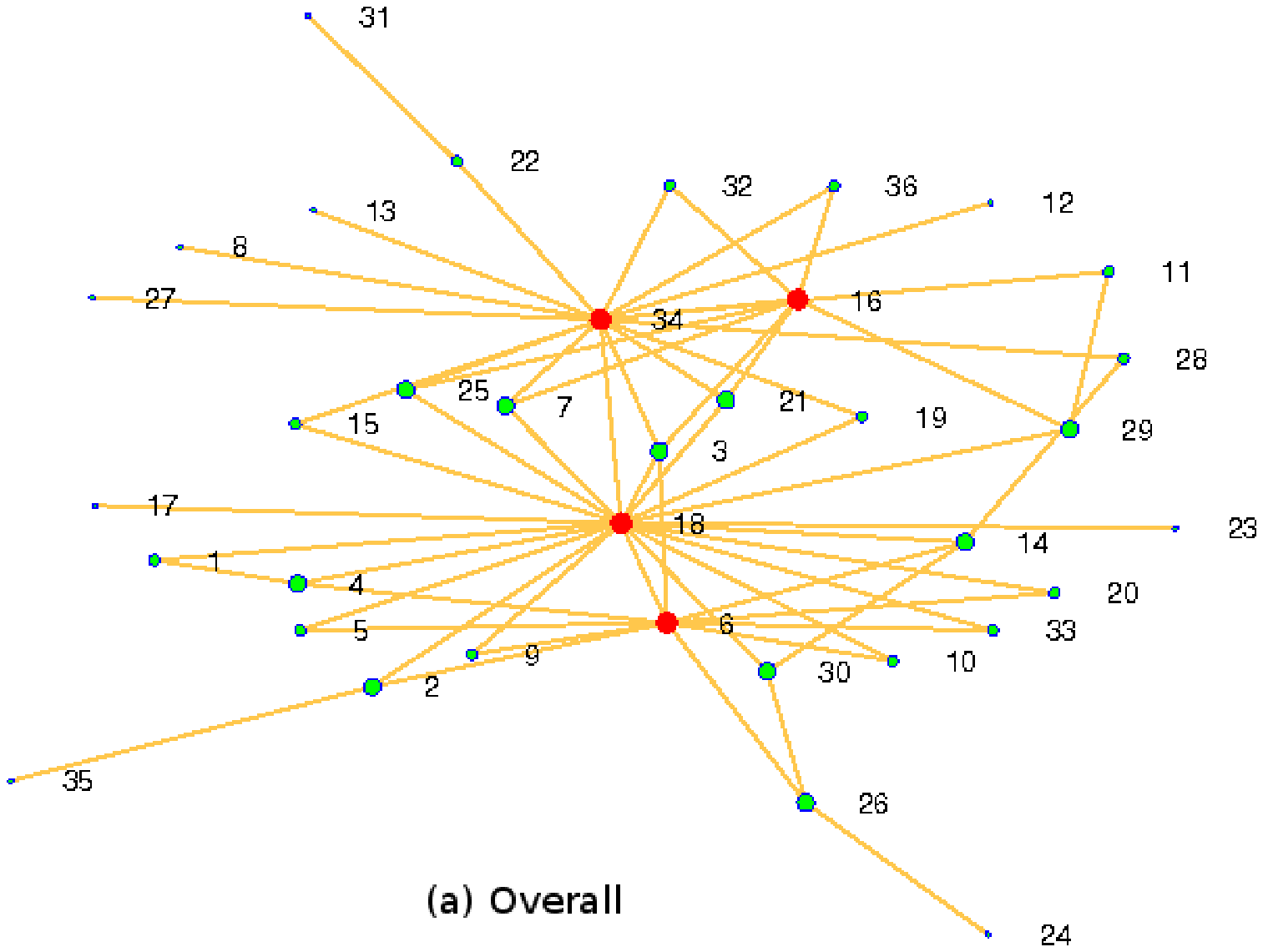} &
\includegraphics[width=4.2cm]{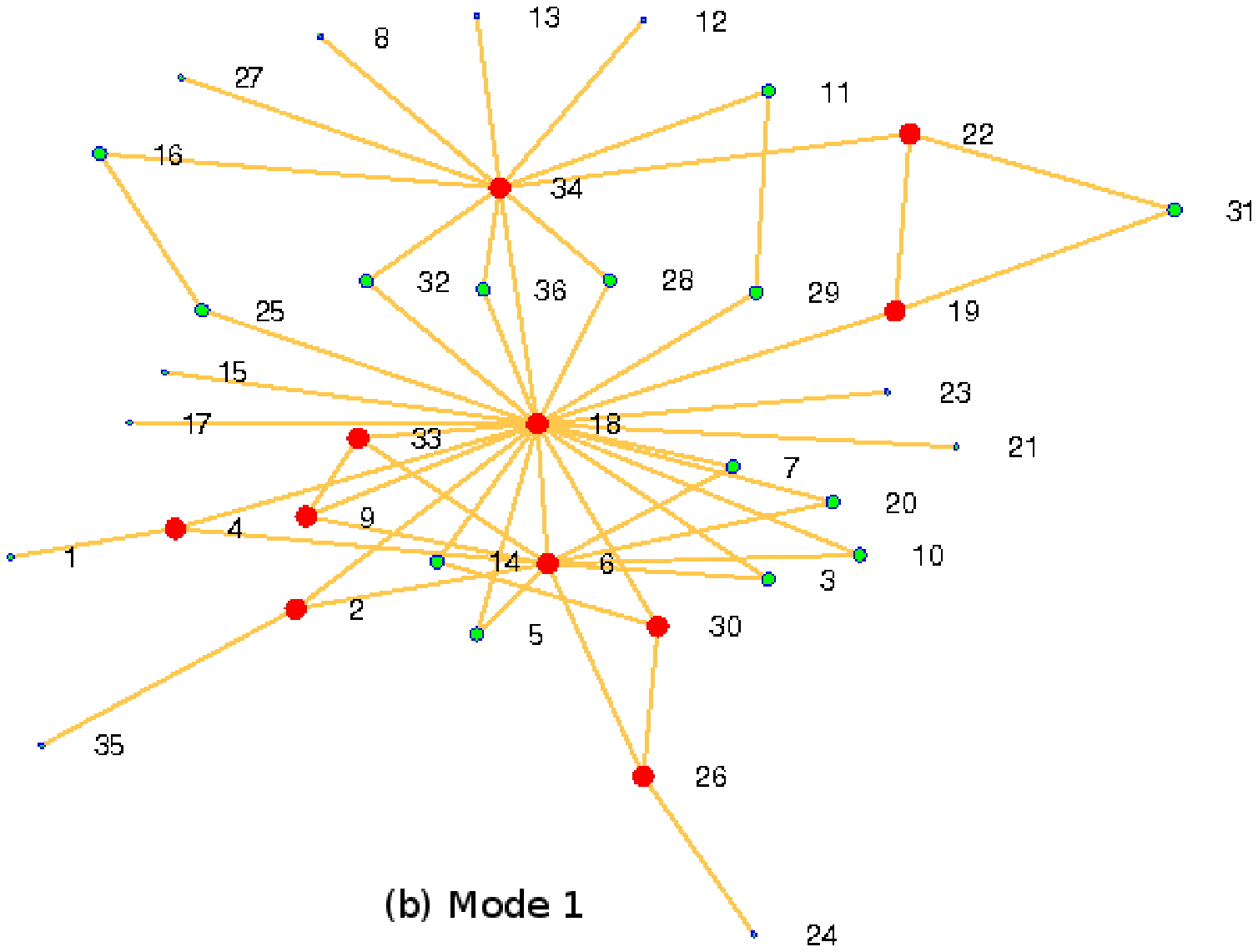} &
\includegraphics[width=4.2cm]{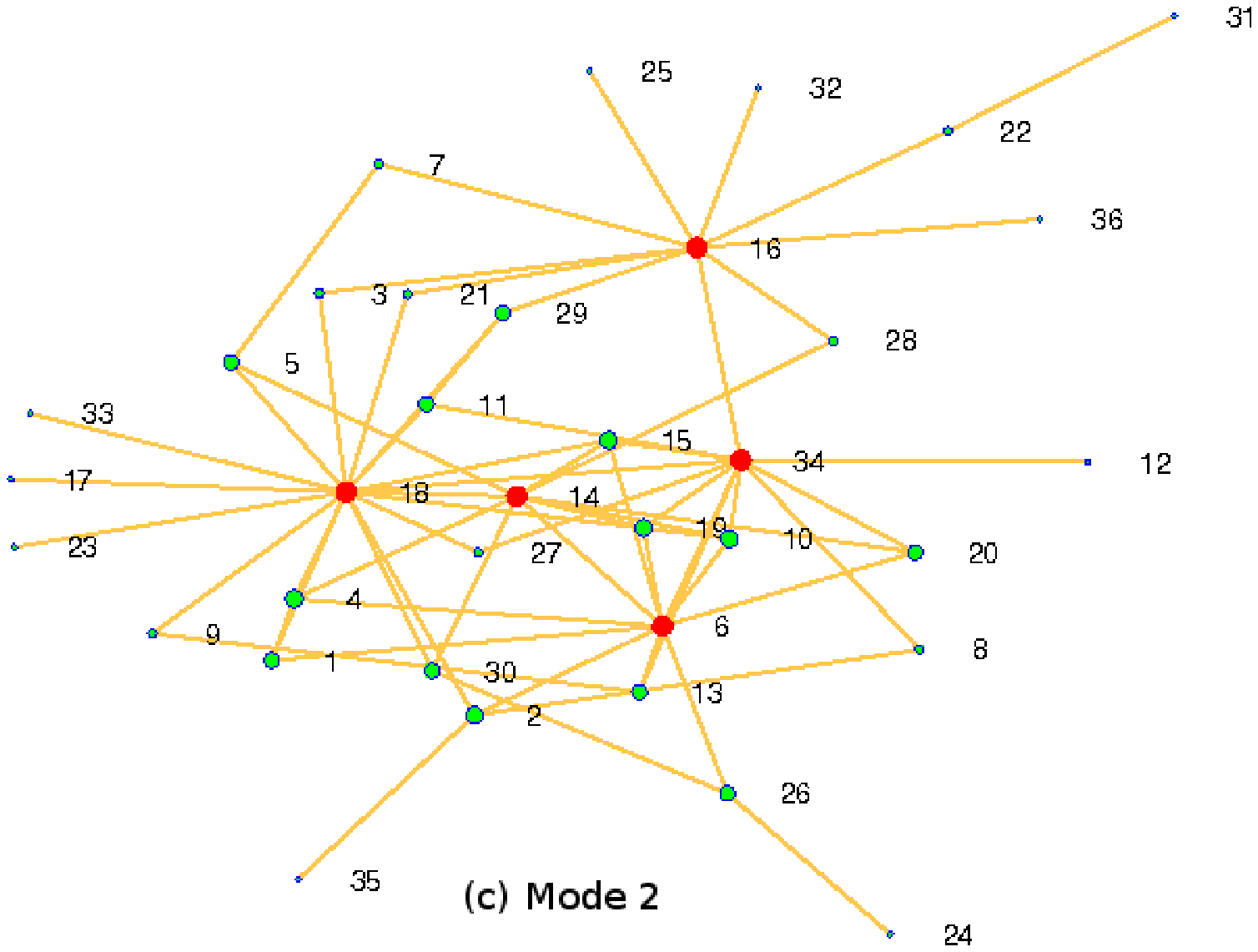} \\
\includegraphics[width=4.2cm]{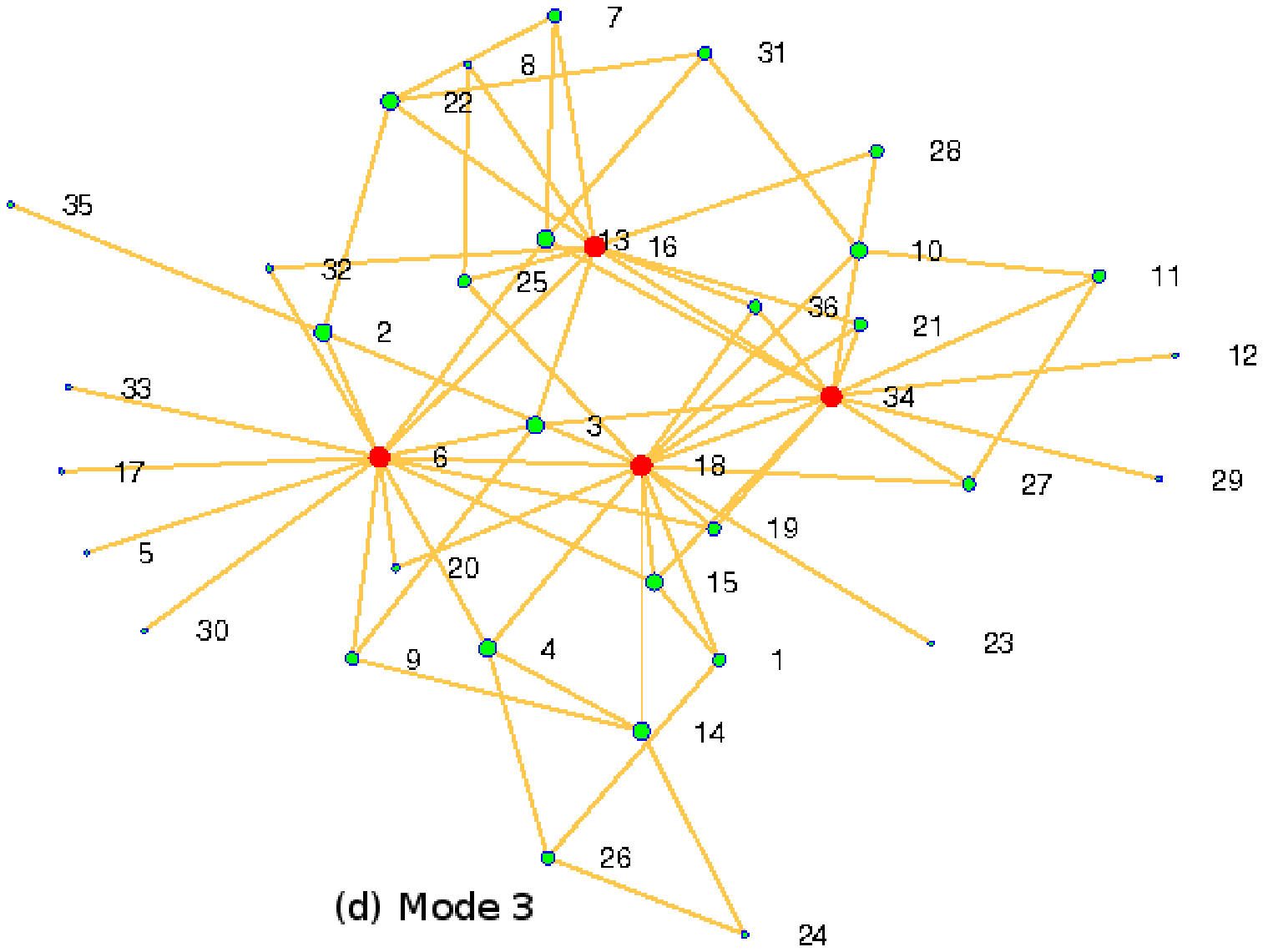} &
\includegraphics[width=4.2cm]{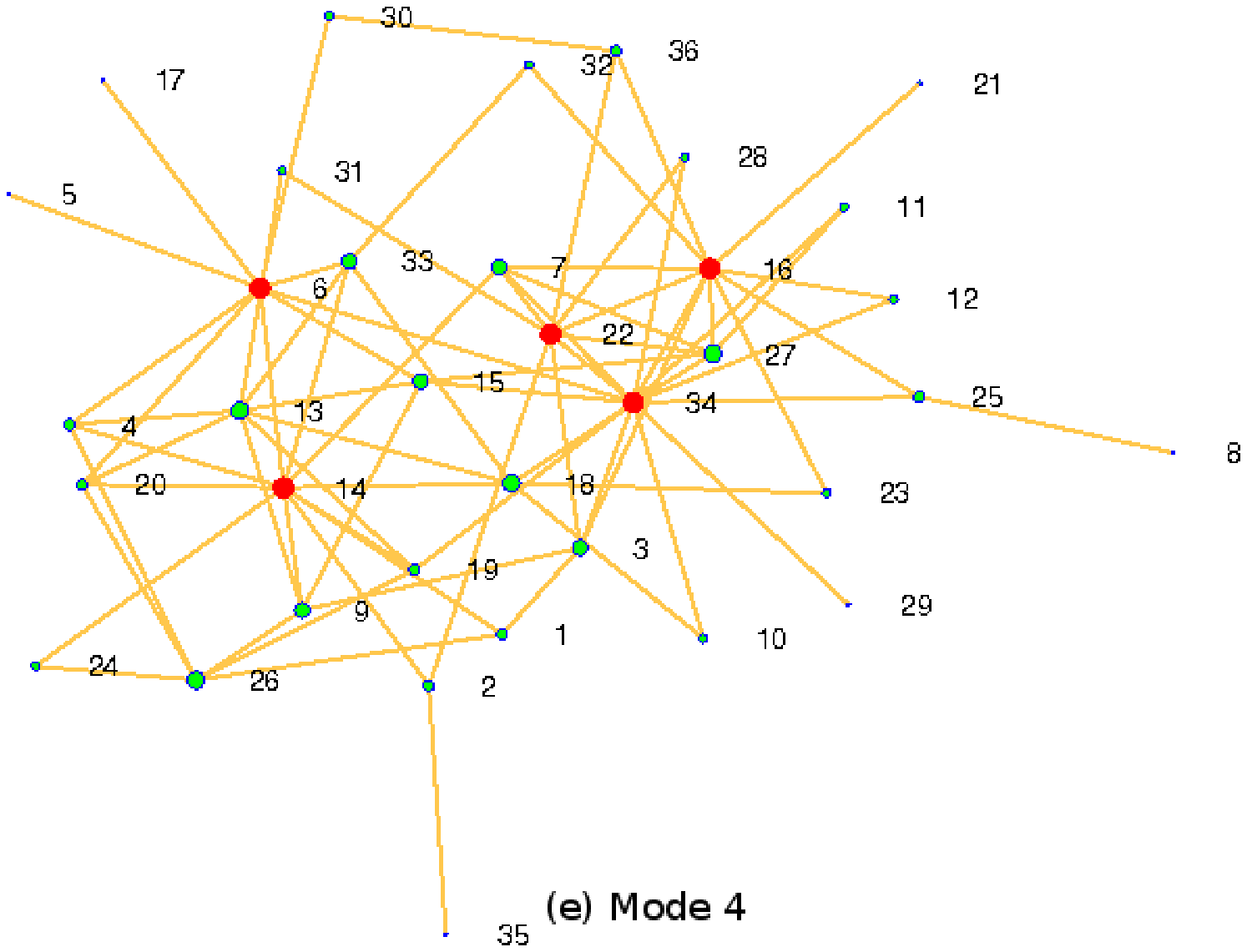} &
\includegraphics[width=4.2cm]{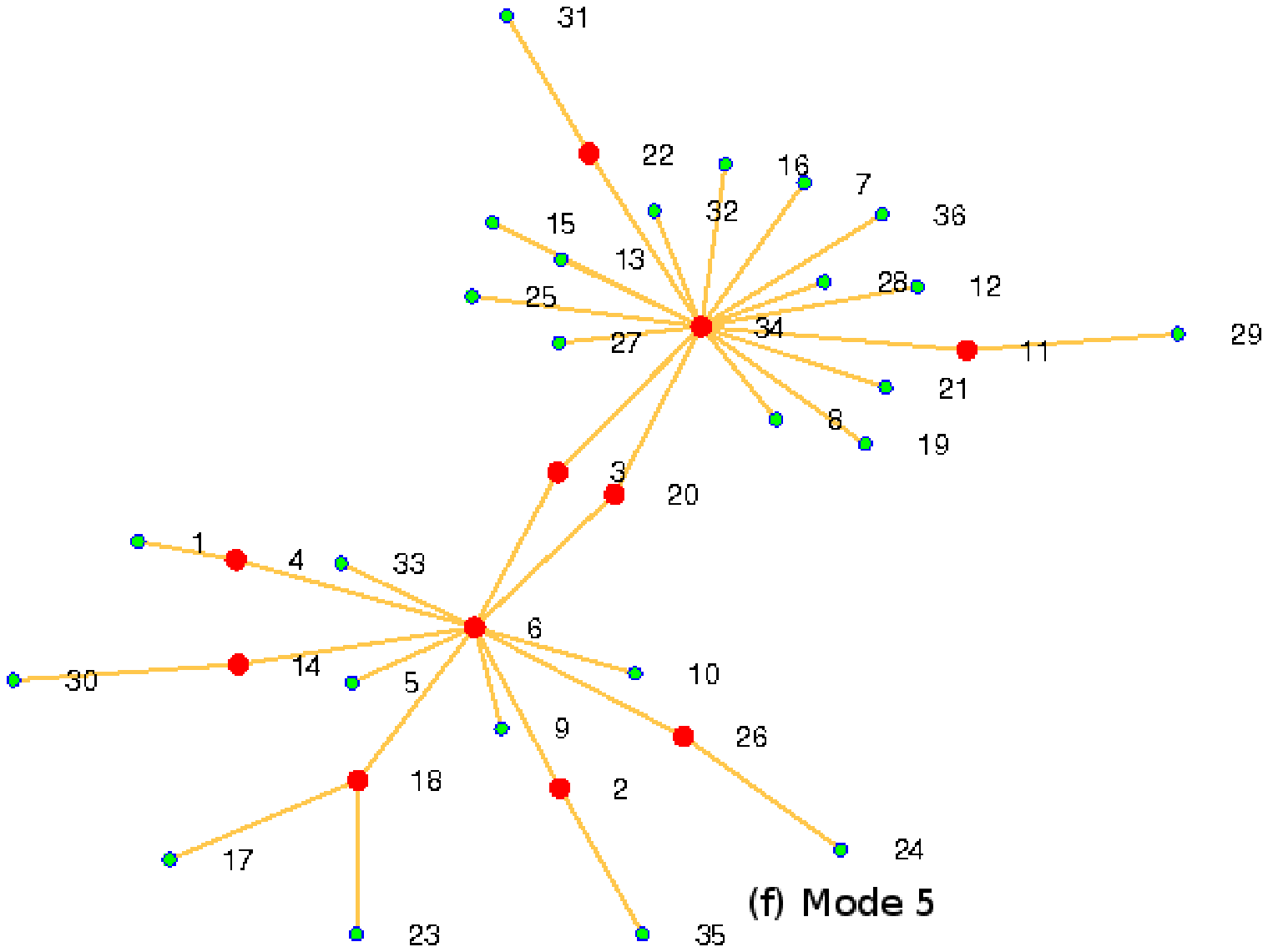}
\end{array}$
\end{center}
\caption{Graph of shortest paths in $\bar{H}$ for overall and 5 modes. (Cambridge data set; the size of a node is proportional to the sum of weights incident on that node)\label{f:camb_summary}}
\end{figure*}

The average graph, $\bar{H}$, for this data set is \emph{represented} in Figure~\ref{f:camb_summary}(a). This representation shows all links in the weighted shortest paths of $\bar{H}$\footnote{We found this to be the clearest means of representing a complete weighted graphs.}. As can be seen the nodes split into two groups as expected. These groups may be represented by a standard dendrogram based on Fiedler vector clustering~\cite{Fiedler1973} as shown in Figure~\ref{f:camb_grouped}. The groups seen here correspond closely to those found on the same data set in~\cite{eiko}.

% \begin{figure}
%  \centering
%    \includegraphics[width=2.4in]{camb_shortest_path_graph.eps}
%  \caption{Graph of shortest paths in $\bar{H}$ (Cambridge data set; the size of a node is proportional to the sum of weights incident on that node). \label{f:camb_shortest_path}}
% \end{figure}
% \begin{figure}
%  \centering
%    \includegraphics[width=3.4in]{camb1.eps}
%  \caption{$\bar{H}$ for the flooding trees \label{f:camb1}}
% \end{figure}
The results so far have examined the average behaviour of the contact network which is interesting in itself. However, by examining the distribution of deviations, $\delta_i$, from the average a more interesting behaviour may be observed. Figure~\ref{f:camb_dist} shows the distribution of $\delta_i$, $i=1\ldots10,000$. As can be seen the distribution is multi-modal; i.e. the underlying process/contact network has different modes of operation. A Gaussian mixture model~\cite{GMM} is used to determine the different modes as shown in Figure~\ref{f:camb_dist}. 5 different modes are identified.
\begin{figure}[tb]
 \centering
   \includegraphics[width=8cm]{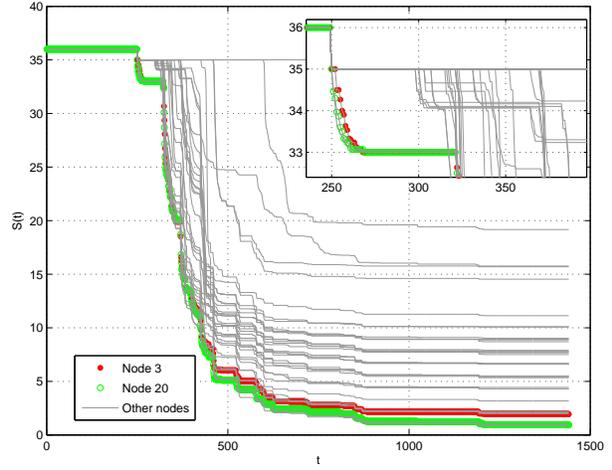}
 \caption{Mean number of nodes susceptible to disease after time $t$ (Root infection starts at time 250. Inset focuses on the start of the infection.). \label{f:seir1}}
\end{figure}
\begin{figure}[b]
 \centering
   \includegraphics[width=8cm]{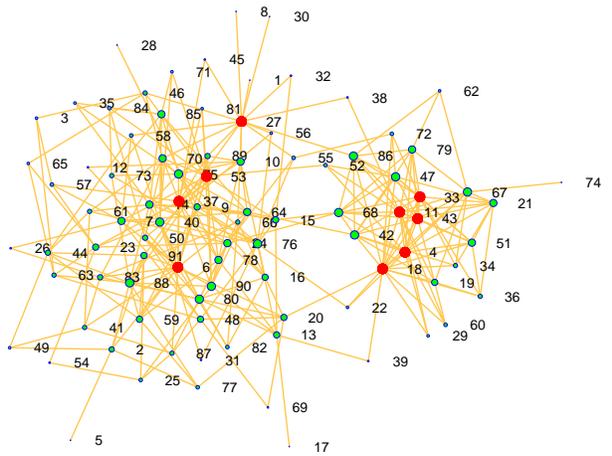}
 \caption{Graph of shortest paths in $\bar{H}$; overall. (MIT). \label{f:MIT_shortest_path_times}}
\end{figure}
\begin{figure}[t]
 \centering
   \includegraphics[width=7cm]{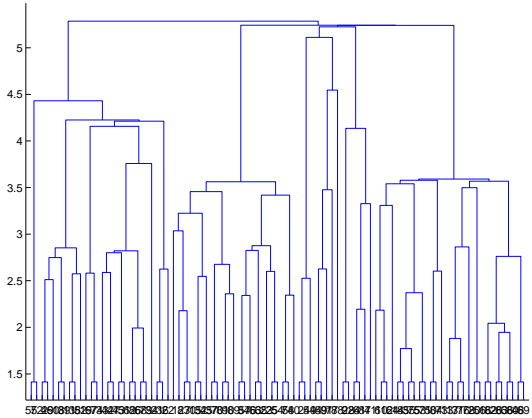}
 \caption{Community based on Fiedler clustering (MIT data set). \label{f:MIT_grouped_clusters}}
\end{figure}
\begin{figure}[t]
 \centering
   \includegraphics[width=6cm]{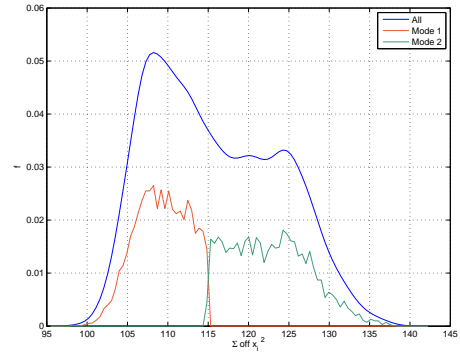}
 \caption{Distribution of $\delta_i$ (MIT).
 \label{f:MIT_dist}}
\vspace{-2mm}
\end{figure}
\begin{figure}[h]
 \centering
   \includegraphics[width=6cm]{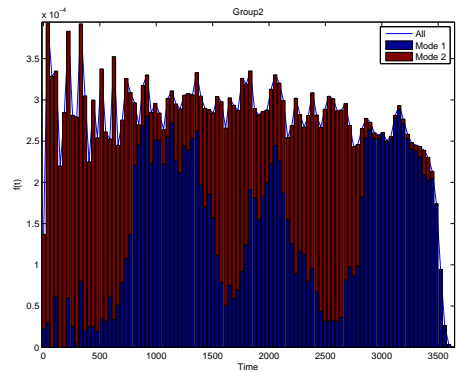}
 \caption{Distribution of times by mode. (MIT) \label{f:MIT_grouped_times}}
\end{figure}
This is particularly useful as it allows the network to be characterized by different modes of behaviour at different times. $\bar{H}$ for the overall data set and for each mode are shown in Figure~\ref{f:camb_summary}. Mode one shows a highly structured network corresponding to the day when the groups are well defined by \emph{class year} (i.e. year 1 and year 2). This structure then becomes less well defined as time moves on. Mode 5 is particularly interesting as there is an obvious bridge formed by nodes 3 and 20. This mode covers the night time and nodes 3 and 20 are possibly staff who interact with the students in the morning\footnote{We cannot be sure as the data has been anonymised.}. Using $\bar{H}$ as an indicator, this implies that a disease spread at this time from nodes 3 and 20 should have the fastest infection rate. Note that mode 1 is still dominant in this period; this mode is essentially being suspended overnight (due to few contacts) with the spanning trees being completed in the morning.

To test the infection rate, an SIR model is constructed\footnote{Probability of infection 0.5; infection time Poisson distributed with mean 80 time steps, 800 mins.} and a disease is spread through the contact network starting at time index 250. The simulation is repeated 30 times for each node and the results bootstrapped to give estimates of the mean number of people susceptible (i.e. those that have not received the disease) at time, $t$, $S(t)$. Figure~\ref{f:seir1} shows the results of these simulations and as can be seen the number of susceptible people falls most rapidly for infections started at nodes 3 and 20, as expected.

\begin{figure*}[htb]
\begin{center}$
\begin{array}{c c c}
\includegraphics[width=7cm]{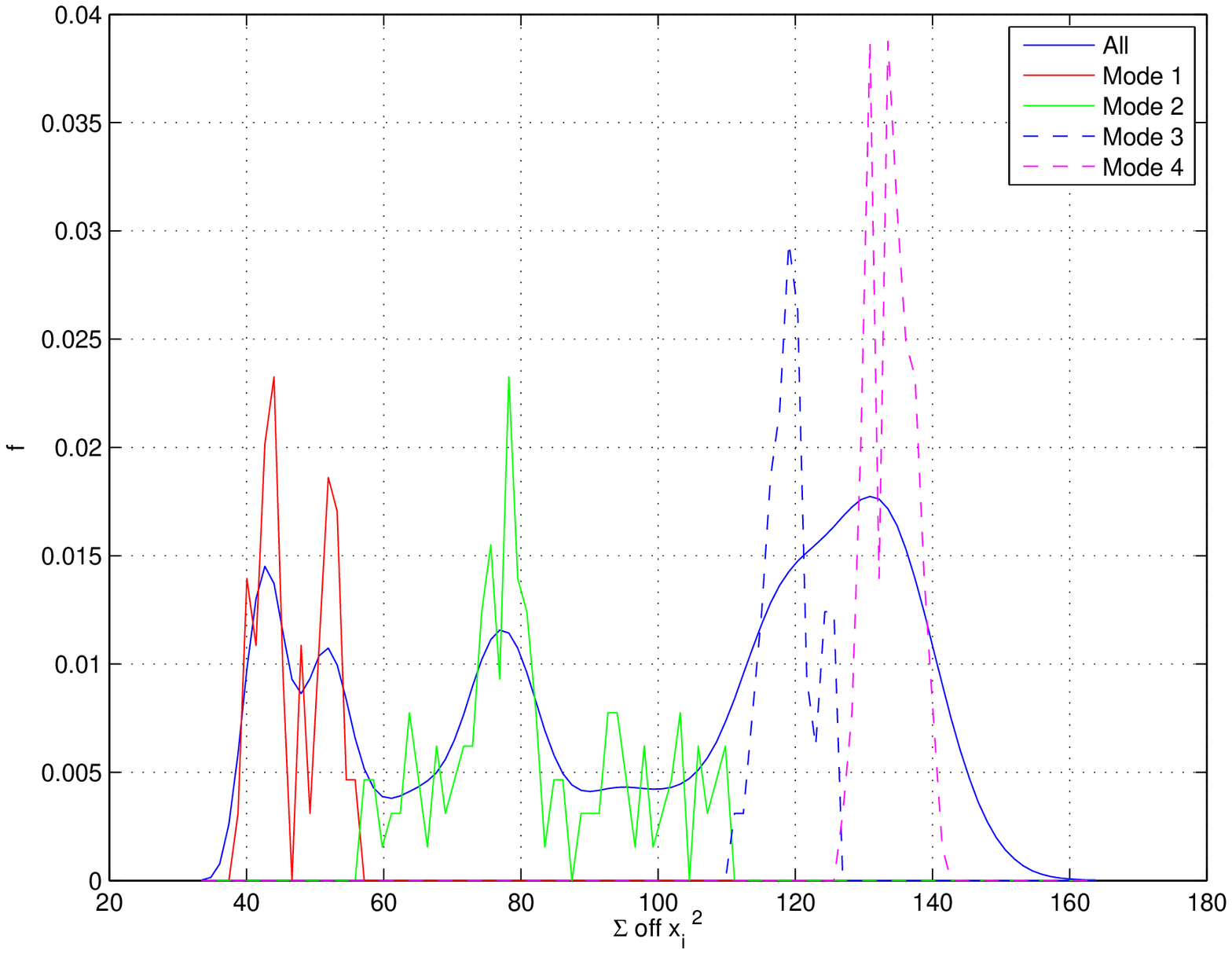} &
\includegraphics[width=7cm]{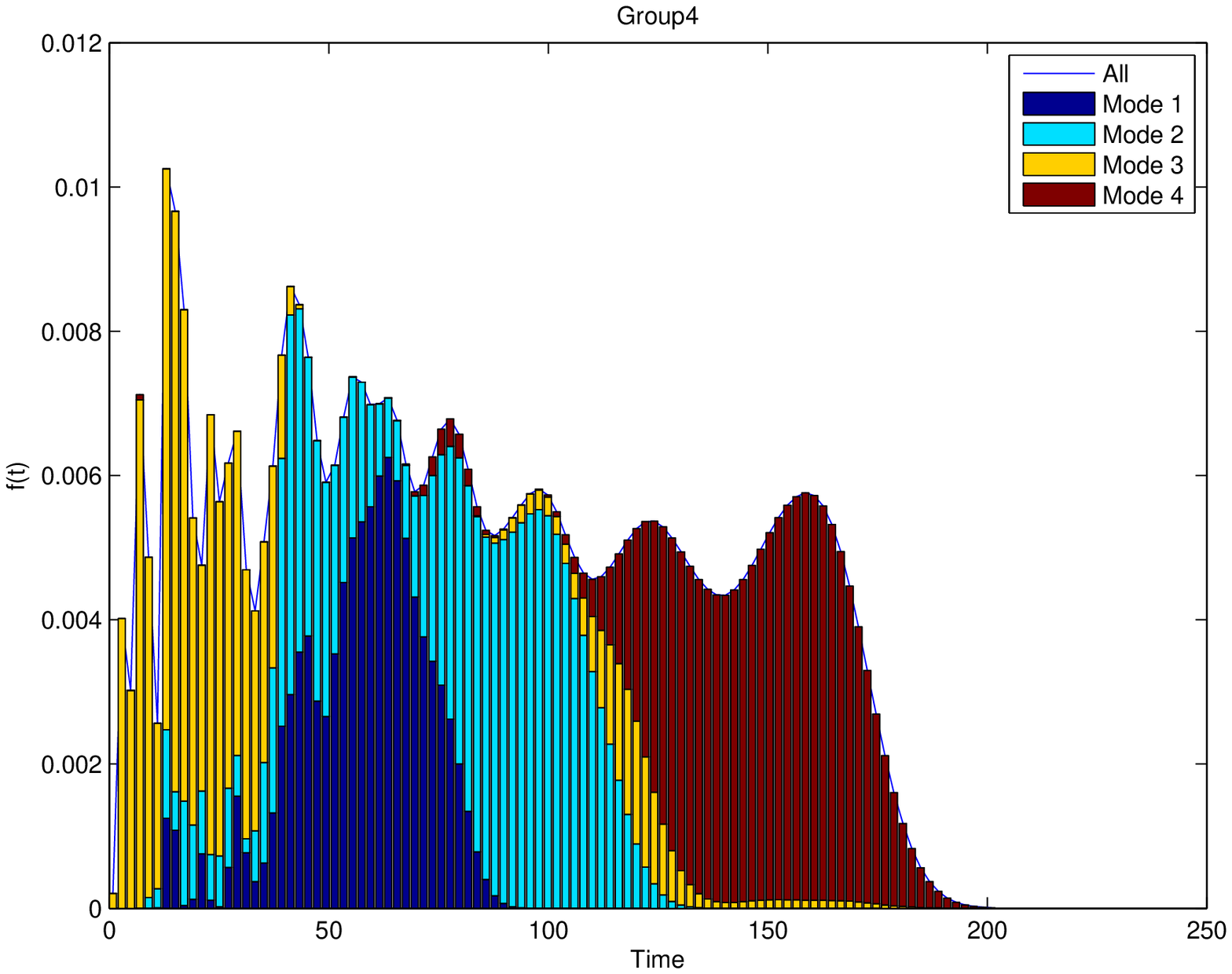}
\end{array}$
\end{center}
\end{figure*}
\begin{figure*}[htb]
\begin{center}$
\begin{array}{c c c}
\includegraphics[width=6.5cm]{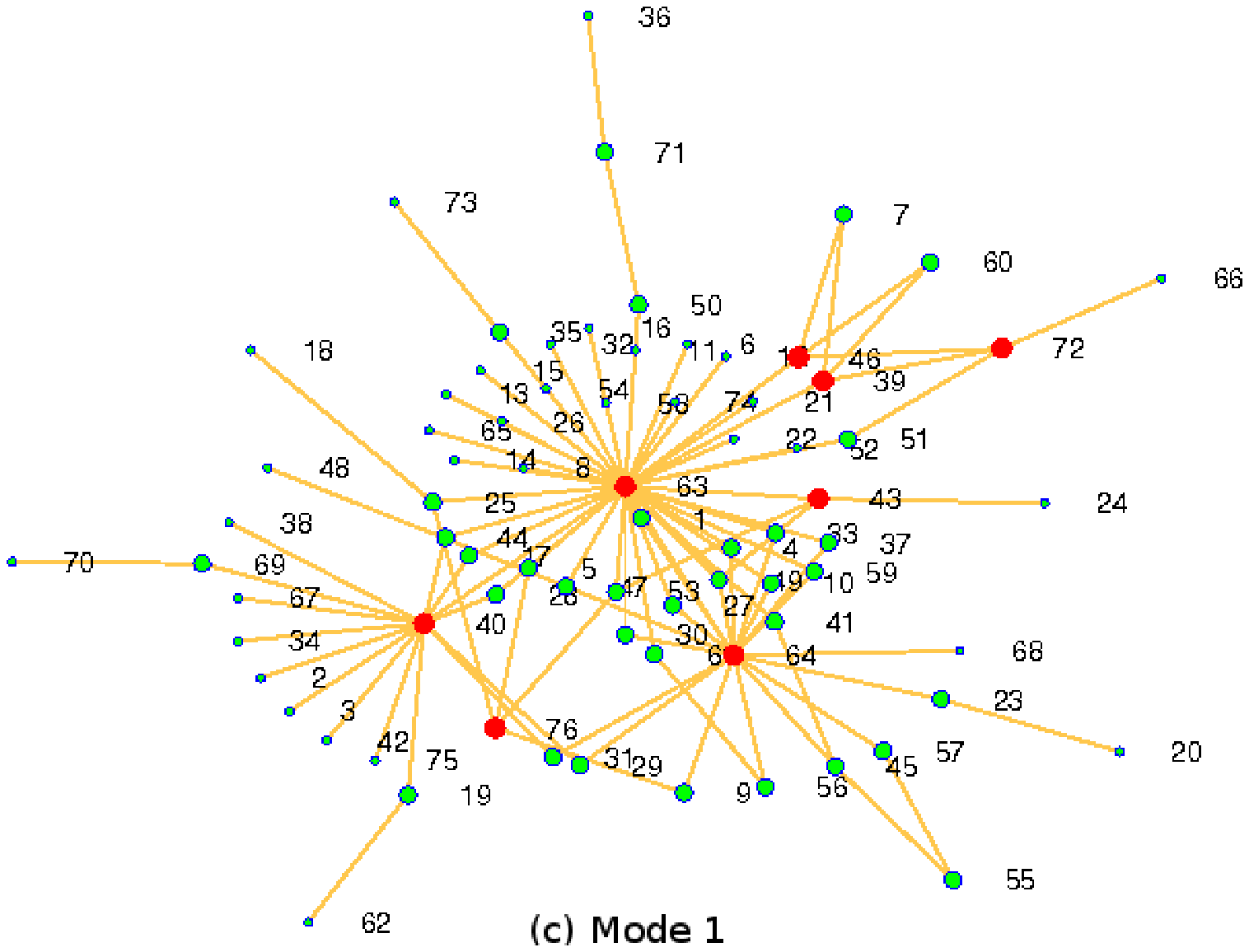} &
\includegraphics[width=6.5cm]{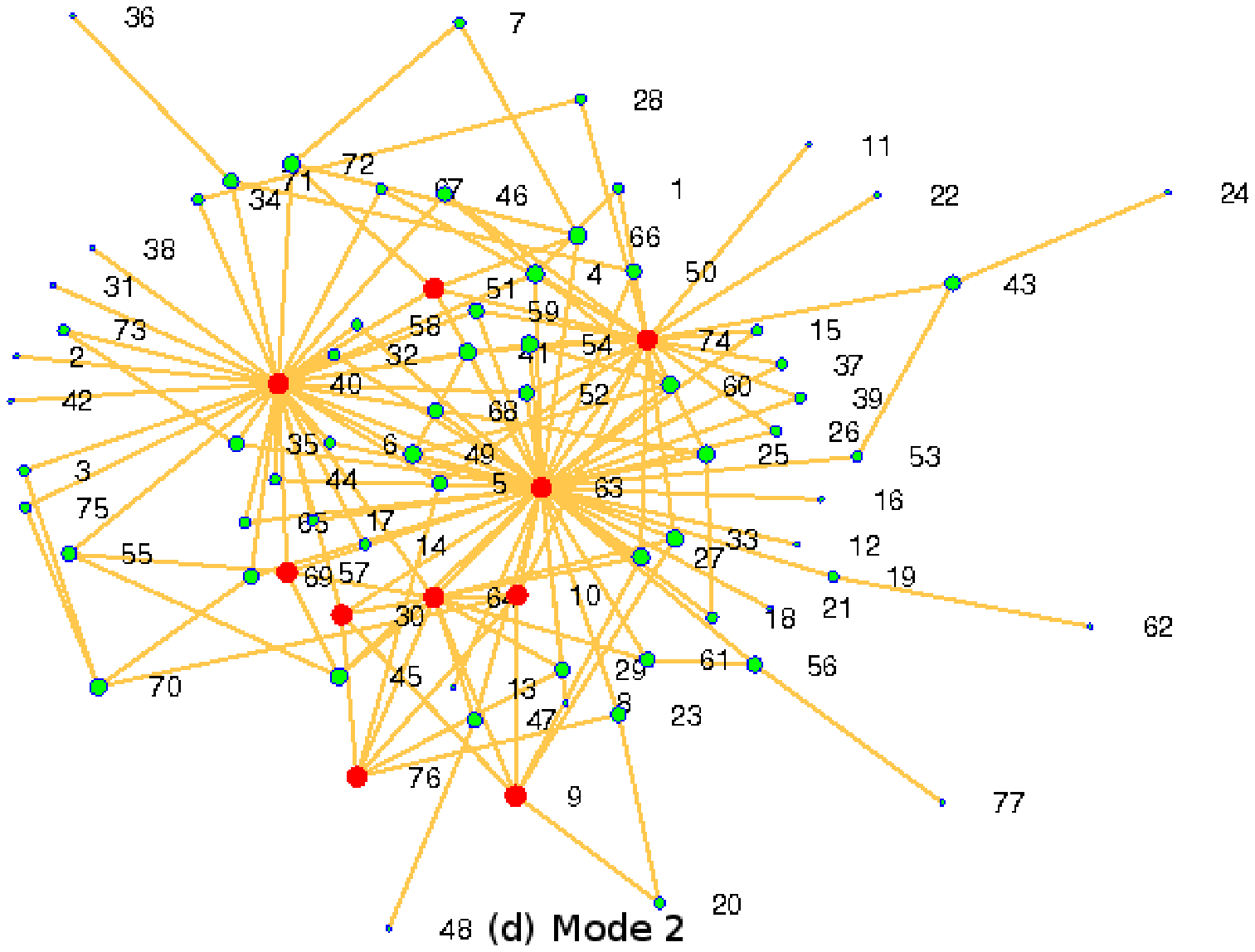} \\
\includegraphics[width=6.5cm]{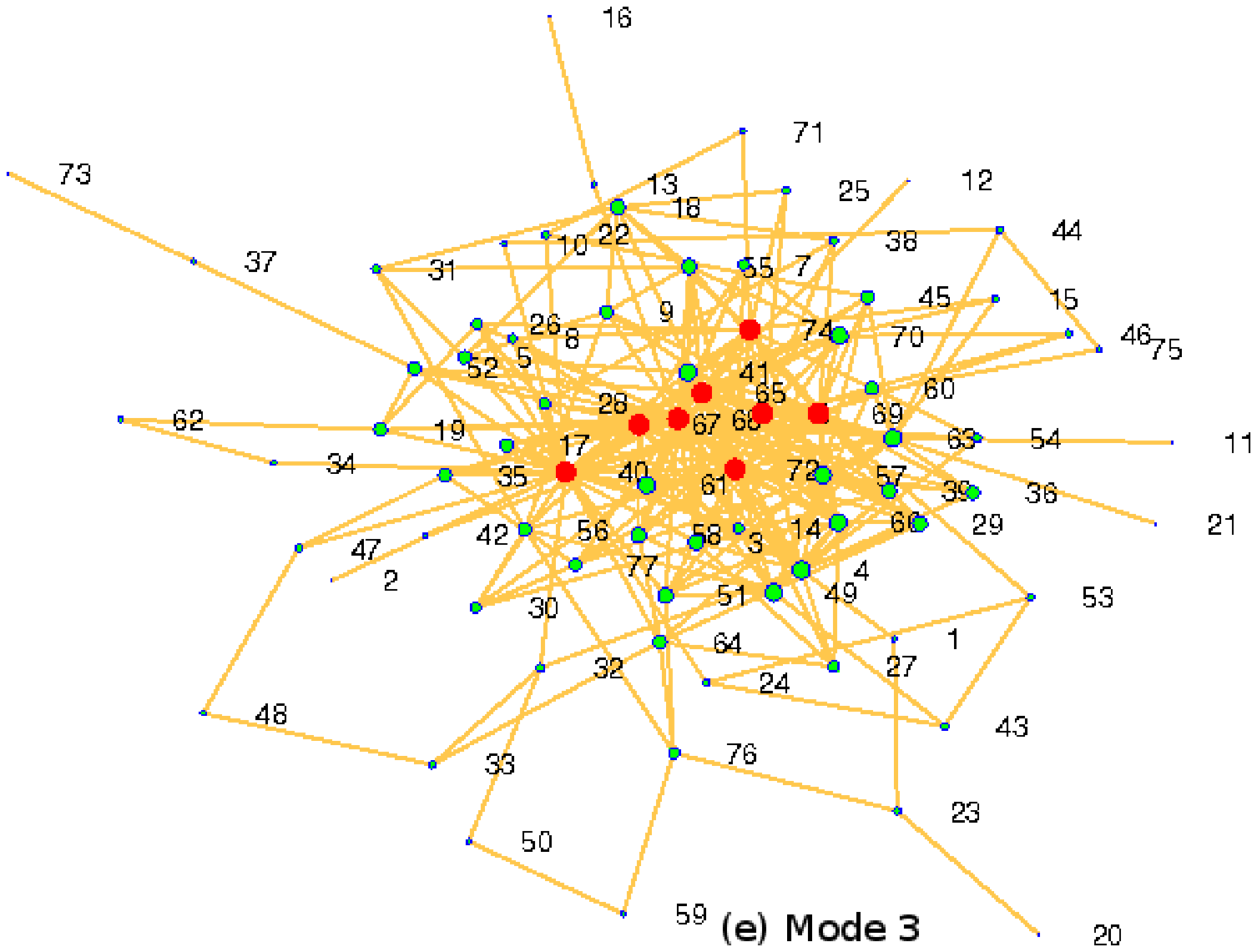} &
\includegraphics[width=6.5cm]{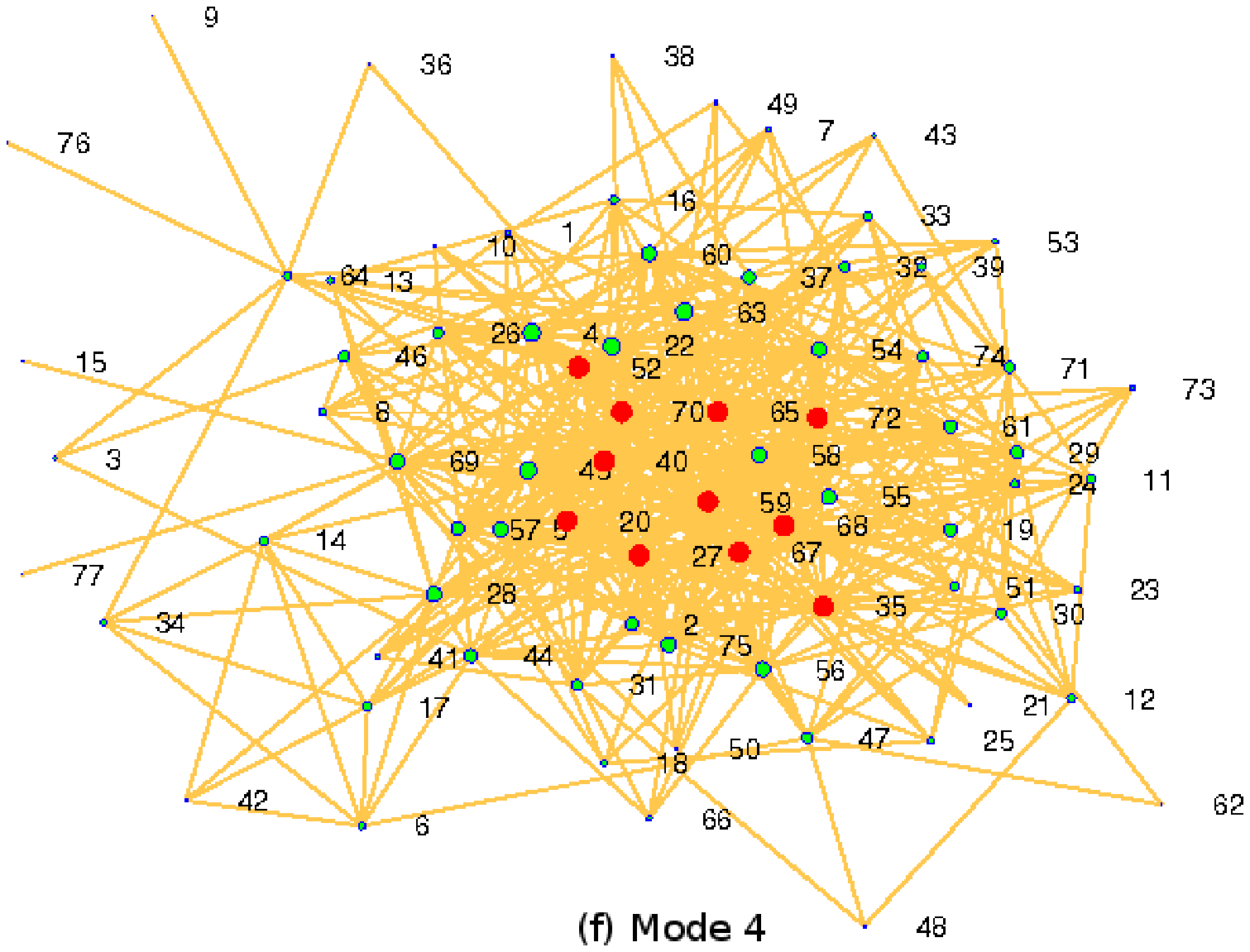}
\end{array}$
\end{center}
\caption{Infocom06 results. (a) Distribution of $\delta_i$. (b) Distribution of times by mode. (c-f) $\bar{H}$
for each mode.\label{f:info_summary}}
\end{figure*}
\subsection{MIT data}
\label{sec:MIT_data}
The results from the MIT data set show a different type of behaviour. There are two main groups (Figure~\ref{f:MIT_shortest_path_times}) the largest of which can be further subdivided into three smaller groups (Figure~\ref{f:MIT_grouped_clusters}).

The distribution of $\delta_i$ is shown in Figure~\ref{f:MIT_dist} and two main modes are identified from this. The MIT data set spans a month of data and recurring patterns emerge from the data as shown in Figure~\ref{f:MIT_grouped_times}. This is particularly interesting as it introduces the concept of being able to forecast the behaviour of a group at regular intervals and design strategies for those specific modes.

\begin{figure*}[htb]
\begin{center}$
\begin{array}{c c c}
\includegraphics[width=5cm]{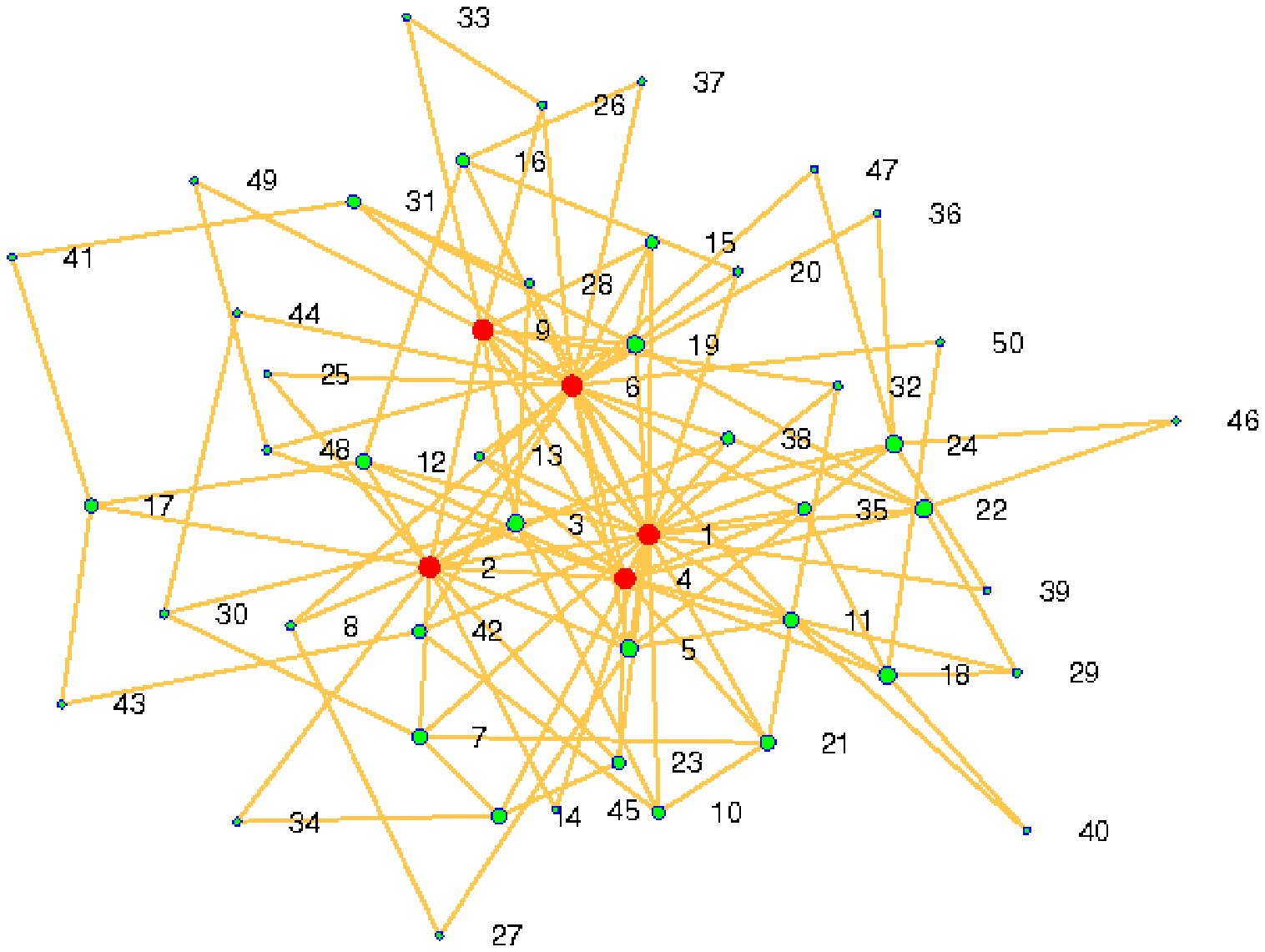} &
\includegraphics[width=5cm]{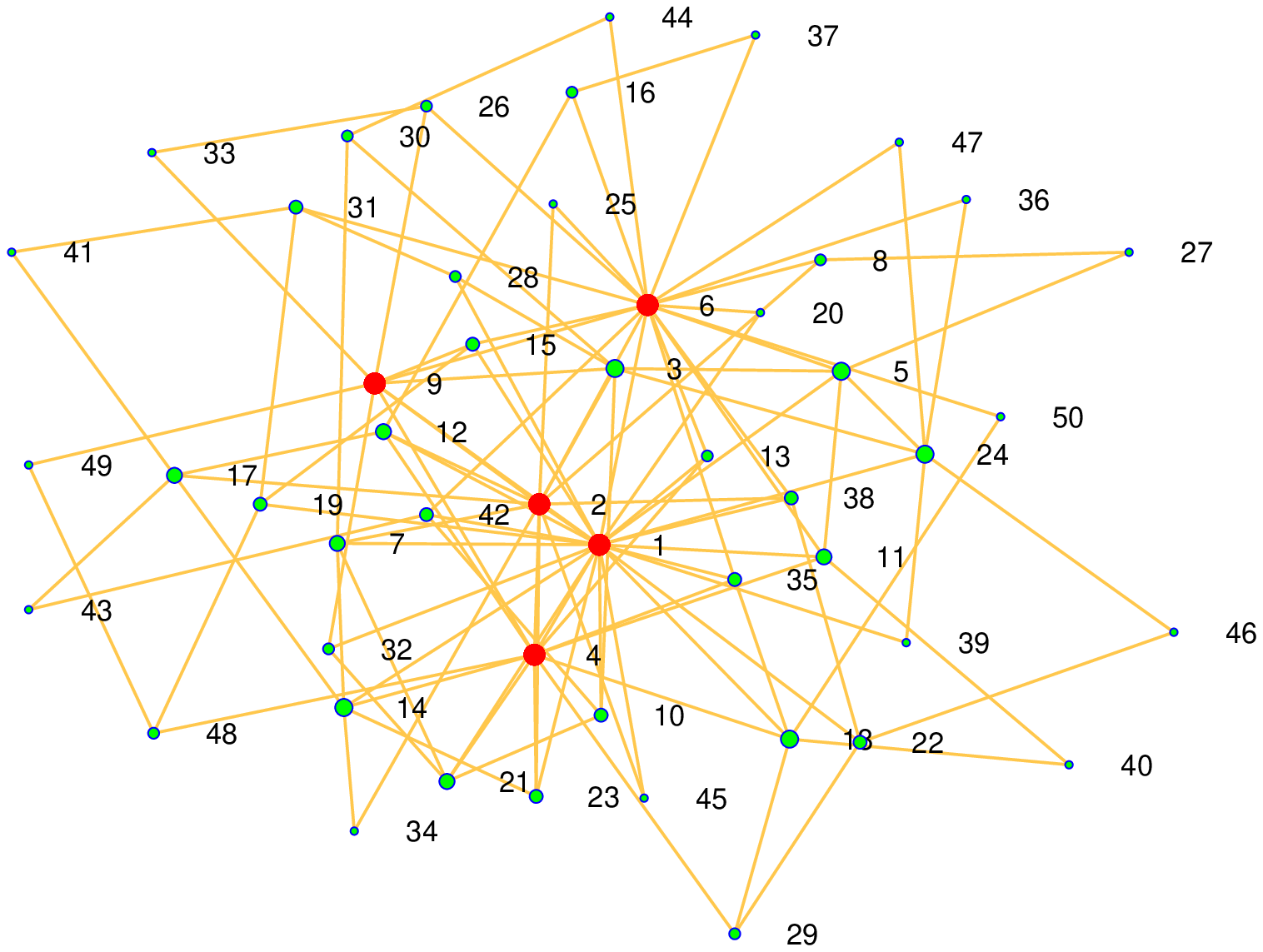}&
\includegraphics[width=5cm]{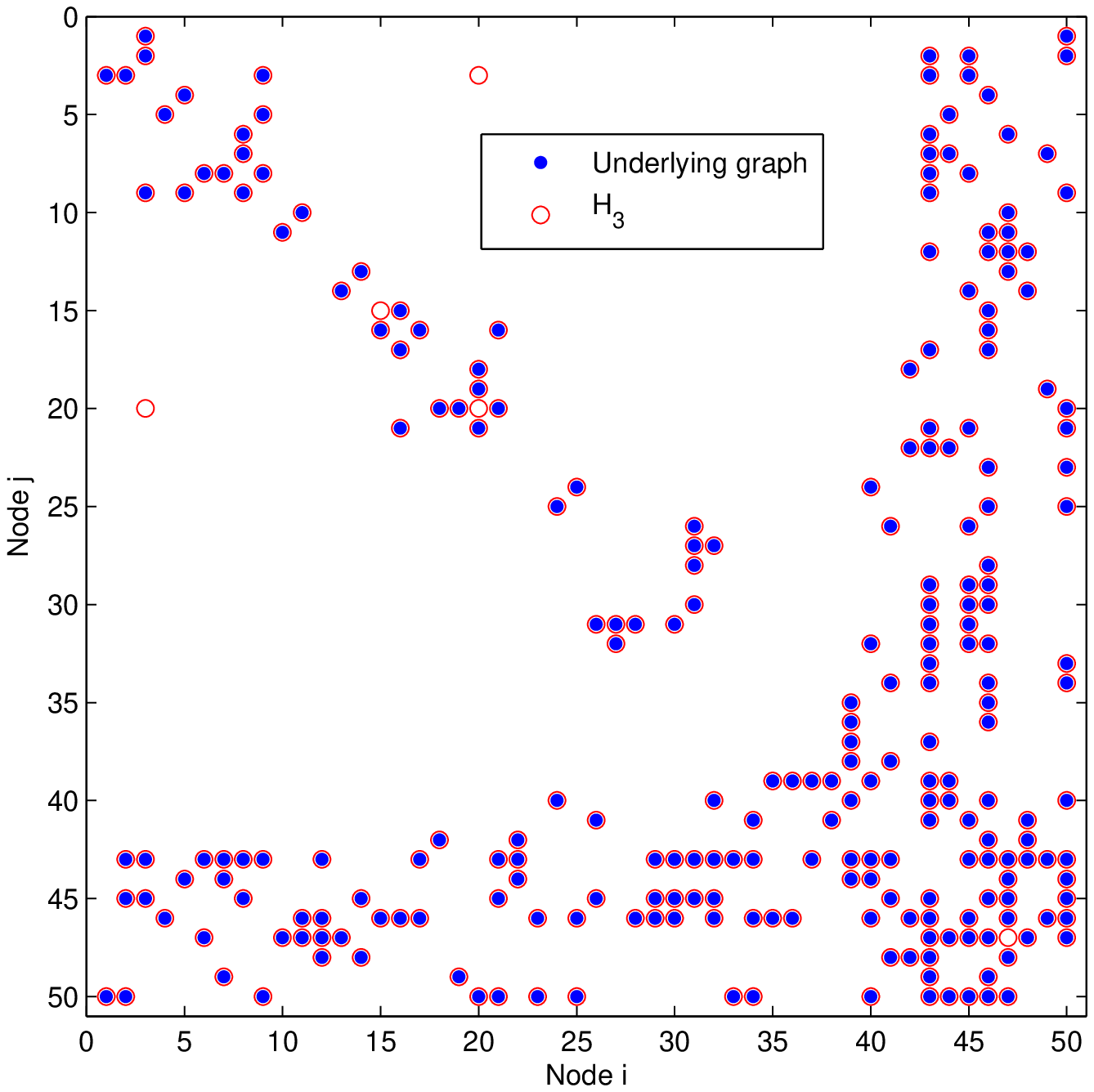}
\end{array}$
\end{center}
\caption{(a) The generating graph for submode 2$^*$. (b) $\bar{H}$ shortest path graph for submode 2$^*$. (c) A
spy plot of the generating graph (.)  and $\bar{H}$ (o)$^{**}$. (Synthetic data) \small{$^*$ The size of a node
is proportional to the sum of weights incident on that node. $^{**}$ The rows and columns have been permuted
using approximate minimum degree to highlight the preferential attachment community structure in the graph.
$\bar{H}$ has been thresholded using a value of 0.1 ($\bar{H}<0.1 \longmapsto 0$).}
\label{f:synth_shortest_mode3}}
\end{figure*}

\subsection{INFOCOM '06 data}
\label{sec:info_data}

The Infocom data is summarised in Figure~\ref{f:info_summary} and follows the behaviour typically expected at a conference. Four modes are identified (Figure~\ref{f:info_summary}(a)) which correspond to four periods in time (Figure~\ref{f:info_summary}(b)). The first mode to occur is mode 3 showing much mixing between the delegates (Figure~\ref{f:info_summary}(e)). This is probably the delegates meeting for coffee before the conference begins. This is then followed by two periods of structured graphs (i.e. presentations; Figures~\ref{f:info_summary}(c,d)) ending with a period of mixing (Figures~\ref{f:info_summary}(f)).
\subsection{Synthetic contact network}
\label{sec:info_data}
\begin{figure}[b]
 \centering
   \includegraphics[width=7cm]{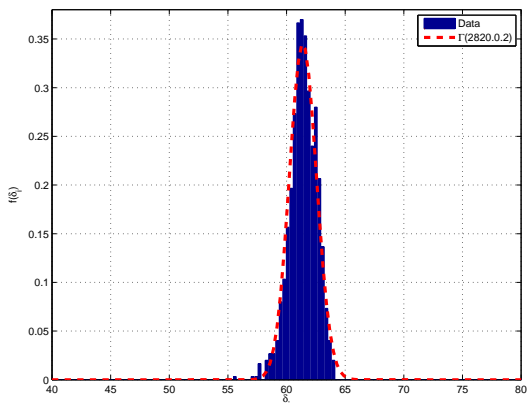}
 \caption{Distribution of $\delta_i$ (Random network). \label{f:rand_dist}}
\end{figure}

The first network created in this section is a purely random contact network in which 5\% of 50 nodes are connected at random in each time step. Figure~\ref{f:rand_dist} shows the distribution of $\delta_i$ for this network is uni-modal as expected; there is only one underlying process. The distribution also follows a $\Gamma$ distribution\footnote{$\delta_i$ is a squared quantity which should follow a similar distribution to a sample variance; i.e. $\delta_i \sim \chi ^2$. The $\Gamma$ distribution is a generalisation of the $\chi ^2$ distribution and so is used.}.

\begin{figure}[htb]
 \centering
   \includegraphics[width=7cm]{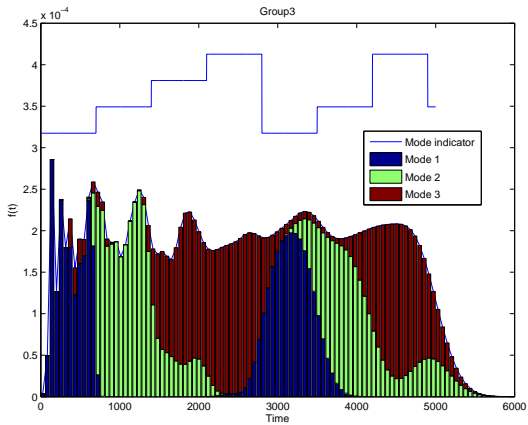}
 \caption{Distribution of times by mode. (Synthetic) \label{f:synth_grouped_times}}
\end{figure}

The second network is more complicated and involves generating four different behaviours for a contact network termed \emph{generators}. A generator consists of a static underlying topology representing a set of possible contacts. These links are transformed into contacts by using a L\'{e}vy walk (as justified in ~\cite{asonam}~\cite{Rhee}); a set of times are generated from a power law distribution and used to demarcate when a contact takes place. The generator used is switched every 700 time units as shown by the mode indicator in Figure~\ref{f:synth_grouped_times}. Specifically, the first generator employs a Waxman topology~\cite{waxman} ($\alpha = 0.5$,$\beta = 0.3$)\footnote{Waxman topology: $p(u\leftrightarrow v)= \alpha e^{-\beta d}$ where $\alpha$ and $\beta$ are parameters of the model. The nodes are distributed randomly on a grid and the distance between them is $d$.}. The second generator is also a Waxman model with $\alpha = 0.7$,$\beta = 0.3$. The third and fourth generators are (GLP) generalised linear preferential topologies based on preferential attachment~\cite{roth:gene}.

\begin{figure}[b]
 \centering
   \includegraphics[width=7cm]{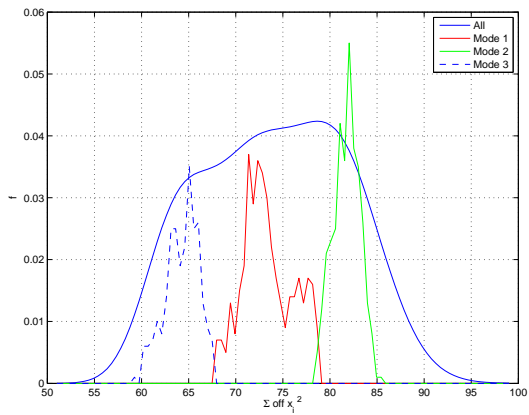}
 \caption{Distribution of $\delta_i$ (Synthetic). \label{f:synth_dist}}
\end{figure}

As can be seen 3 modes are detected in the data (Figure~\ref{f:synth_dist}). These correspond with the generator times for 2 of the modes (Figure~\ref{f:synth_grouped_times}). However, mode 3 incorporates both generator 3 and generator 4. This occurs as generator 3 and 4 are quite similar (both based on GLP).

\begin{figure}[htb]
 \centering
   \includegraphics[width=7cm]{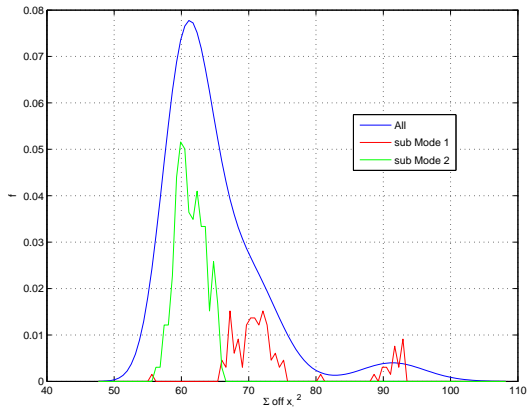}
 \caption{Distribution of $\delta_i$; submode. (Synthetic). \label{f:synth_dist_sub}}
\end{figure}

The samples in mode 3 may be examined separately using JD to produce the \emph{submodes} seen in Figure~\ref{f:synth_dist_sub} occurring at the times seen in Figure~\ref{f:synth_dist_sub}. As can be seen these submodes are generators 3 and 4. Thus the algorithm has successfully recovered the modes in the data. At this point we make a note on the transition between the modes. It is interesting that this transition is not crisp even though the switching between modes is. This is because a spanning tree may begin in one mode but the message may end in the next mode.

\begin{figure}[htb]
 \centering
   \includegraphics[width=7cm]{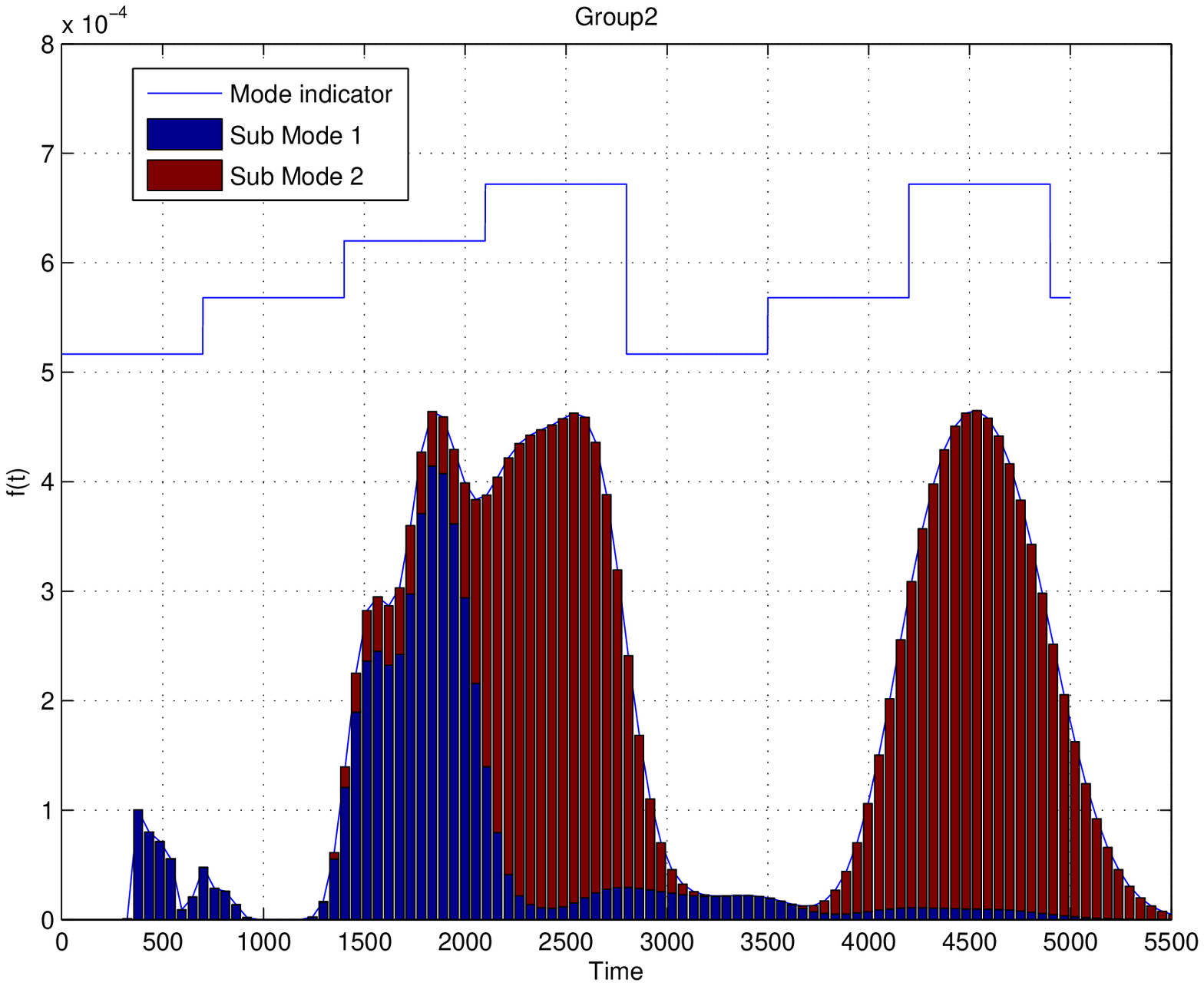}
 \caption{Distribution of times by mode; submode. (Synthetic) \label{f:synth_grouped_times_sub}}
\end{figure}

% %%%%%%%%%%%%%%%%%%%%%%%%%%%%%%%%%%%%%%%%%%%%
%
% % \begin{figure}
% %  \centering
% %    \includegraphics[width=3.4in]{camb2.eps}
% %  \caption{Hierarchy of the groups using Fiedler clustering. \label{f:cambex}}
% % \end{figure}
%
% - time constraint.
% - the effect of a sub-tree - i.e. zero entries in the matrices.
% - how to construct an average graph
%
% Synthetic network 1 - sample bias.
%
% Synthetic network 2 - preferred routes.
%
% AS level graph centrality.
%
% Social Network Analysis.
%
%

%
% \begin{figure}
%  \centering
%    \includegraphics[width=2.4in]{camb_shortest_path_graph2.eps}
%  \caption{Graph of shortest paths in $\bar{H}$ mode 5. (Cambridge data set). \label{f:camb_shortest_path_5_times}}
% \end{figure}
%

% infocomm results.

%%%%%%%%%%%%%%%%%%%%%%%%%%%%%%%%%%%%%%%%%%%
%%%%%%%%%%%%%%%%%%%%%%%%%%%%%%%%%%%%%%%%%%%
%%%%%%%%%%%%%%%%%%%%%%%%%%%%%%%%%%%%%%%%%%%
\section{Conclusions}
\label{sec:conclusion}
%%%%%%%%%%%%%%%%%%%%%%%%%%%%%%%%%%%%%%%%%%%
%%%%%%%%%%%%%%%%%%%%%%%%%%%%%%%%%%%%%%%%%%%
%%%%%%%%%%%%%%%%%%%%%%%%%%%%%%%%%%%%%%%%%%%

This paper presented a method for extracting different modes of operation for contact networks. In the
real-world contact networks examined, several interesting features where extracted including detection of a
bridge in the Cambridge data set. The MIT data set in contrast, showed a repetitive behaviour which is useful
for prediction of network behaviour; for example in advance of an infection. The INFOCOM data set clearly
showed the behaviour typical of a conference. In producing an average graph based on samples of a network, the
order of contacts has been preserved and in addition the correlation between contacts has been preserved. For
example aggregation based purely on counting the number of times a link is present does not take into account
the fact that links may typically be present \emph{together}; i.e. the time based correlation between links. By
using spanning trees the methodology takes advantage of a sampling mechanism present in many real-world
networks; it might not be possible to record all contacts but it is often possible to flood a message in a
network and record the paths taken. It is hoped that in future this technique will aid in the design of time
specific algorithms for time dependent networks.

\section*{Acknowledgment}
The research is part funded by the EU grants for the Recognition project, FP7-ICT-257756 and the EPSRC DDEPI
Project, EP/H003959. We would like to thank the members of Systems Research Group, University of Cambridge, for
their comments and suggestions.

\bibliographystyle{abbrv}

\bibliography{topology}

\end{document}